\documentclass[twocolumn,english,unsortedaddress,notitlepage,longbibliography]{revtex4-1}
\usepackage[T1]{fontenc}
\usepackage[latin9]{inputenc}
\usepackage{color}
\usepackage{array}
\usepackage{multirow}
\usepackage{amsmath}
\usepackage{amssymb}
\usepackage{graphicx}

\makeatletter

\newcommand{\lyxmathsym}[1]{\ifmmode\begingroup\def\b@ld{bold}
  \text{\ifx\math@version\b@ld\bfseries\fi#1}\endgroup\else#1\fi}

\newcommand{\angstrom}{\textup{\AA}}

\providecommand{\tabularnewline}{\\}

\makeatother
\usepackage{babel}
\begin{document}
\title{Transition to metallization in warm dense helium-hydrogen mixtures
using stochastic density functional theory within the Kubo-Greenwood
formalism}

\author{Yael Cytter}
\affiliation{Fritz Haber Center for Molecular Dynamics and Institute of Chemistry,The Hebrew University of Jerusalem, Jerusalem 9190401, Israel}

\author{Eran Rabani}
\email{eran.rabani@berkeley.edu}
\affiliation{Department of Chemistry, University of California and Materials Science
	Division, Lawrence Berkeley National Laboratory, Berkeley, California
	94720, U.S.A.}
\affiliation{The Raymond and Beverly Sackler Center for Computational Molecular
	and Materials Science, Tel Aviv University, Tel Aviv, Israel 69978}

\author{Daniel Neuhauser}
\email{dxn@chem.ucla.edu}
\affiliation{Department of Chemistry, University of California at Los Angeles,
	CA-90095 USA}

\author{Martin Preising}

\author{Ronald Redmer}
\email{ronald.redmer@uni-rostock.de}
\affiliation{Institute of Physics, University of Rostock, A.-Einstein-Str. 23,18059 Rostock, Germany}
\author{Roi Baer}
\email{roi.baer@huji.ac.il}
\affiliation{Fritz Haber Center for Molecular Dynamics and Institute of Chemistry,The Hebrew University of Jerusalem, Jerusalem 9190401, Israel}

\begin{abstract}
The Kubo-Greenwood (KG) formula is often used in conjunction with
Kohn-Sham (KS) density functional theory (DFT) to compute the optical
conductivity, particularly for warm dense mater. For applying the
KG formula, all KS eigenstates and eigenvalues up to an energy cutoff
are required and thus the approach becomes expensive, especially for
high temperatures and large systems, scaling cubically with both system
size and temperature. Here, we develop an approach to calculate the
KS conductivity within the stochastic DFT (sDFT) framework, which
requires knowledge only of the KS Hamiltonian but not its eigenstates
and values. We show that the computational effort associated with
the method scales linearly with system size and reduces in proportion
to the temperature unlike the cubic increase with traditional deterministic
approaches. In addition, we find that the method allows an accurate
description of the entire spectrum, including the high-frequency range,
unlike the deterministic method which is compelled to introduce a
high-frequency cut-off due to memory and computational time constraints.
We apply the method to helium-hydrogen mixtures in the warm dense
matter regime at temperatures of $\sim60\text{kK}$ and find that
the system displays two conductivity phases, where a transition from
non-metal to metal occurs when hydrogen atoms constitute $\sim0.3$
of the total atoms in the system.
\end{abstract}
\maketitle

\section{\label{sec:Introduction}Introduction}

The state of warm dense matter (WDM) is characterized by high atomic
density, similar to conventional condensed matter systems, and elevated
temperatures of several electron volts ($1\text{eV}\approx10^{4}K$).
This is an intermediate regime bridging plasma physics and condensed
matter physics for which equations of state (EOS) and other properties
are of interest. One example appears in the study of hydrogen-helium
mixtures under extreme conditions, where the EOS \citep{militzer2013equation},
phase separation and physical properties, such as conductivity \citep{lorenzen2011metallization}
and miscibility \citep{schottler2018abinitio} can be used to explain
the luminosity and gravitational moments of planets such as Jupiter
and other gas giants, as well as their formation and evolution characteristics
\citep{Stevenson1975,nettelmann2008abinitio,Guillot1999}. Generally,
EOS and properties are calculated for various materials using first-principle
methods, specifically the Kohn-Sham density functional theory (KS-DFT)
at finite temperatures \citep{Silvestrelli1996,Mattsson1997,pozzo2012thermal,witte2018observations},
often showing good agreement with experiments \citep{witte2018observations,Holst2008,Preising2018}.
Within the KS-DFT framework, WDM conductivity is often obtained by
using the Kubo-Greenwood (KG) formalism \citep{kubo1957statisticalmechanical,mazevet2010calculations,holst2011electronic,desjarlais2002electrical}
with good results when compared to experiment. The KS-DFT and the
KG electrical conductivity equation when applied to WDM requires large
computational effort which increases dramatically with temperature
and system size, because of the need to construct and propagate all
the occupied KS eigenstates, as well as a sufficient number of unoccupied
states, the number of which grows as $T^{3}$, where $T$ is the temperature
\citep{cytter2018stochastic}).

Recently, stochastic DFT (sDFT) approaches that circumvent the computational
difficulties mentioned above have been developed \citep{Baer2013,Neuhauser2014a,cytter2014metropolis,cytter2018stochastic,fabian2018stochastic,chen2019overlapped}
for ground/thermal state calculations. These have also served as a
basis for developing time-dependent methodologies for description
of materials properties \citep{Gao2015,Neuhauser2017,hernandez2018first,takeshita2017stochastic}.
It was shown that sDFT is especially useful for EOS calculations in
the WDM regime since it involves a computational effort that scales
as $T^{-1}$ \citep{cytter2018stochastic}.

In this paper, we develop an approach for calculating the KG conductivity
within the framework of sDFT. The main advantage of the approach is
that it does not require any knowledge of the occupied or empty KS
orbitals. We show and benchmark a stochastic method to sample the
KG conductivity. We then use the method to study the conductivity
in hydrogen-helium mixtures. Our approach is similar to previously
developed stochastic conductivity approaches \citep{Wang1994g,Baer2004c,Iitaka1997}
but differs in essential implementation details and is unique in its
combination with sDFT calculations.

In the paper, we present the development of the stochastic KG (sKG)
method and provide important implementation details, as well as demonstrations
of the methods validity and a discussion in the statistical errors
and scaling in Sec.~\ref{sec:Method}. In~Sec.~\ref{sec:Mixed-systems}
the sDFT-sKG method is applied to the study of the conductivity of
mixtures helium and hydrogen in the warm dense matter regime, targeting
metallization and beyond-linear-mixing effects.

\section{Method\label{sec:Method}}

\subsection{Time-dependent linear response\label{subsec:Time-dependent-linear-response}}

The time-dependent expectation value of a many-body observable $\hat{B}$
\emph{after }an impulsive perturbation is applied through the observable
\textbf{$\hat{A}$} to a system at time $t=0$ (usually assumed in
thermal equilibrium) is given, in the linear-response regime, as the
following correlation function \citep{kubo1957statisticalmechanical,kubo1966thefluctuationdissipation}:
$C_{AB}\left(t\right)=i\theta\left(t\right)\text{Tr}\left[\rho\left(\beta,\mu\right)\left[\hat{A},\hat{B}\left(t\right)\right]\right]$
where $\hat{B}\left(t\right)=e^{i\hat{H}t/\hbar}\hat{B}e^{-i\hat{H}t/\hbar}$,
$\hat{H}$ is the unperturbed Hamiltonian and $\theta\left(t\right)$
is the Heaviside function imposing causality. The expectation values
are performed with respect to the many-body thermal density $\rho\left(\beta,\mu\right)=Z\left(\beta,\mu\right)^{-1}e^{-\beta\left(\hat{H}-\mu\hat{N}\right)}$
where $Z\left(\beta,\mu\right)$ is the partition function at chemical
potential $\mu$ and inverse temperature $\beta=\frac{1}{k_{B}T}$,
$k_{B}$ being the Boltzmann constant.

One of the important applications of linear-response theory is the
prediction of the frequency-dependent conductivity 
\begin{equation}
\sigma(\omega)=\frac{2\pi e^{2}}{\Omega m_{e}^{2}\hbar}\frac{\Im\left(\tilde{C}_{PP}(\omega)\right)}{\omega}\label{eq:SigmaResponseRelation}
\end{equation}
where $\Omega$ is the volume of the simulation cell and $\tilde{C}_{PP}(\omega)$
is the Fourier transform of the momentum-momentum correlation function,

\begin{equation}
\tilde{C}_{PP}\left(\omega\right)=\int_{0}^{\infty}C_{PP}\left(t\right)e^{-i\omega t}e^{-\frac{1}{2}\eta^{2}t^{2}}dt,\label{eq:Cpp}
\end{equation}
and $\eta$ is a small real parameter. In the limit $\omega\to0$
L'hopital's rule can be used to assess the DC conductivity:
\begin{equation}
\sigma\left(0\right)=\frac{2\pi e^{2}}{\Omega m_{e}^{2}\hbar}\lim_{\omega\to0}\frac{\partial\Im\tilde{C}_{PP}\left(\omega\right)}{\partial\omega}.\label{eq:sigma(0)}
\end{equation}
For non-interacting particles, with a single-particle Hamiltonian
$\hat{h}$, having eigenvalues $\varepsilon_{n}$ and eigenstates
$\left|n\right\rangle $, $n=1,2,...$, the correlation function reduces
to the following expression:

\begin{equation}
C_{AB}\left(t\right)=-2\theta\left(t\right)\Im\text{Tr}\left[f_{FD}\left(\hat{h}\right)\hat{a}\left(1-f_{FD}\left(\hat{h}\right)\right)\hat{b}\left(t\right)\right]\label{eq:correlationAB}
\end{equation}
where $\hat{a}$, \textbf{$\hat{b}$} are the single-particle perturbing
and observed operators, respectively, \textbf{$\hat{b}\left(t\right)=e^{i\hat{h}t}\hat{b}e^{-i\hat{h}t}$}
and 
\begin{equation}
f_{FD}\left(\hat{h}\right)\equiv\frac{1}{1+e^{\beta\left(\hat{h}-\mu\right)}}
\end{equation}
is the Fermi-Dirac distribution. Combining Eq.~\ref{eq:correlationAB}
and Eq.~\ref{eq:Cpp} and taking the formal limit $\eta\rightarrow0$
gives the Kubo-Greenwood (KG) conductivity \citep{kubo1957statisticalmechanical,greenwood1958theboltzmann}:
\begin{equation}
\sigma\left(\omega\right)=\frac{2\pi e^{2}}{\Omega m_{e}^{2}\hbar\omega}\sum_{m,n}^{N_{g}}f_{mn}\left|p_{mn}\right|^{2}\delta\left(\omega-\varepsilon_{nm}/\hbar\right),\label{eq:conduct_deterministic}
\end{equation}
where $N_{g}$ is the number of grid points, $f_{mn}\equiv f_{FD}\left(\varepsilon_{m}\right)-f_{FD}\left(\varepsilon_{n}\right)$,
$\varepsilon_{nm}=\varepsilon_{n}-\varepsilon_{m}$ and $p_{nm}=\left\langle n\left|\hat{p}\right|m\right\rangle $.
For practical reasons, the summation over the occupied and unoccupied
states is determined according to an energy cutoff and as a result
the conductivity spectrum can be calculated only up to a corresponding
frequency cutoff.

\subsection{\label{subsec:Stochastic-response-function}Stochastic calculation
of the response function}

To calculate the KG conductivity in a stochastic manner the stochastic
trace formula \citep{Hutchinson1990} can be used to estimate the
trace in Eq.~(\ref{eq:correlationAB}). However, we found that a
smaller statistical noise can be obtained if the stochastic trace
is applied to following equivalent but more symmetrical expression:
\begin{equation}
C_{PP}\left(t\right)=-2\theta\left(t\right)\Im\text{Tr}\left[\sqrt{f_{FD}}\hat{p}\left(1-f_{FD}\right)\hat{p}\left(t\right)\sqrt{f_{FD}}\right].\label{eq:symResponse}
\end{equation}
To apply the stochastic trace formula, we define a set of stochastic
orbitals $\chi$, represented on the grid such that $\left\langle \boldsymbol{r}_{g}|\chi_{i}\right\rangle =\left(\delta x\right)^{-3/2}e^{i\theta_{g}^{i}}$,
where $\theta_{g}\in\left[0,2\pi\right]$ is a random phase and $\delta x$
is the grid spacing.The stochastic expression for $C_{PP}\left(t\right)$
is given by:

\begin{align}
C_{PP}\left(t\right) & =-2\theta\left(t\right)\text{E}\left\{ \Im\left\langle \xi\left|\hat{p}\left(1-f_{FD}\left(\hat{h}\right)\right)e^{i\hat{h}t}\hat{p}e^{-i\hat{h}t}\right|\xi\right\rangle \right\} ,\label{eq:final-sKG-2}
\end{align}

\noindent \begin{flushleft}
where $\left|\xi\right\rangle =\sqrt{f_{FD}\left(\hat{h}\right)}\left|\chi\right\rangle $
and $\text{E}\left\{ ...\right\} $ designates an expectation value.
\par\end{flushleft}

The procedure consists of the following schematic steps:
\begin{enumerate}
\item Set: $n=0$, $\left|\eta_{j}\right\rangle =\left|\xi_{j}\right\rangle $,
$\left|\zeta_{j}\right\rangle =\left(1-f_{FD}\left(\hat{h}\right)\right)\hat{p}\left|\xi_{j}\right\rangle $,
and the time-step $\Delta t=\frac{\pi}{\Delta E}$, where $\Delta E=E_{max}-E_{min}$
and $E_{max}$ ($E_{min})$ is the maximal (minimal) eigenvalue of
$\hat{h}$ (the condition is required to avoid aliasing). The time
step determines the cutoff frequency of the spectrum and $N_{ts}=\frac{\Delta E}{\Delta\omega}=\frac{\pi}{\Delta\omega\Delta t}$
is the total number of time steps for achieving a spectral resolution
of $\Delta\omega$.
\item Calculate: $C_{PP}^{j}\left(n\Delta t\right)=-2\Im\left\langle \zeta_{j}\left|\hat{p}\right|\eta_{j}\right\rangle $.
\item \label{enu:Set--,}Set $n=n+1$ , $\left|\eta_{j}\right\rangle =e^{-i\hat{h}\Delta t}\left|\eta_{j}\right\rangle $,
$\left|\zeta_{j}\right\rangle =e^{-i\hat{h}\Delta t}\left|\zeta_{j}\right\rangle $.
\item Go to 2 and repeat until $n=N_{ts}$.
\item The response function is then averaged over $I_{\sigma}$ (the number
of stochastic orbitals), yielding $C_{PP}\left(n\Delta t\right)\approx\frac{1}{I_{\sigma}}\sum_{j=1}^{I_{\sigma}}C_{PP}^{j}\left(n\Delta t\right)$.
This response function is then discrete-Fourier transformed and used
to obtain the frequency-dependent conductivity (Eq.~(\ref{eq:SigmaResponseRelation})).
\end{enumerate}
The process is easily parallelized, since each element $C_{PP}^{j}\left(n\Delta t\right)$
is calculated independently before averaging in the final step. In
our calculations we do not use the above procedure directly because
using Chebyshev expansions for the evolution operator, we found a
way to expedite the calculation as described in Sec.~\ref{subsec:Algorithmic-implementation-and}.

The stochastic-KG (sKG) procedure forms a post processing step after
a sDFT calculation \citep{Baer2013,cytter2018stochastic} which provides
the self consistent KS Hamiltonian $\hat{h}$. The stochastic calculation
requires two sets of stochastic orbitals, one set is used to perform
sDFT calculation with which $\hat{h}$ is determined, this set will
be denoted ``sDFT-os'' and a second set, used in the sKG calculation
to determine the conductivity is denoted ``sKG-os''. The KS wave
functions are expanded using plane waves although the non-local part
of the pseudopotentials are implemented using a real-space grid, for
achieving high efficiency. For all the stochastic calculations in
this paper, we used the local density approximation (LDA) \citep{Perdew1992a}
and Troullier-Martins norm-conserving pseudopotentials \citep{Troullier1991}
within the Kleinman-Bylander representation \citep{Kleinman1982}.

\subsection{\label{sec:Validation}Validation of the method}

\begin{figure}
\includegraphics[width=0.9\columnwidth]{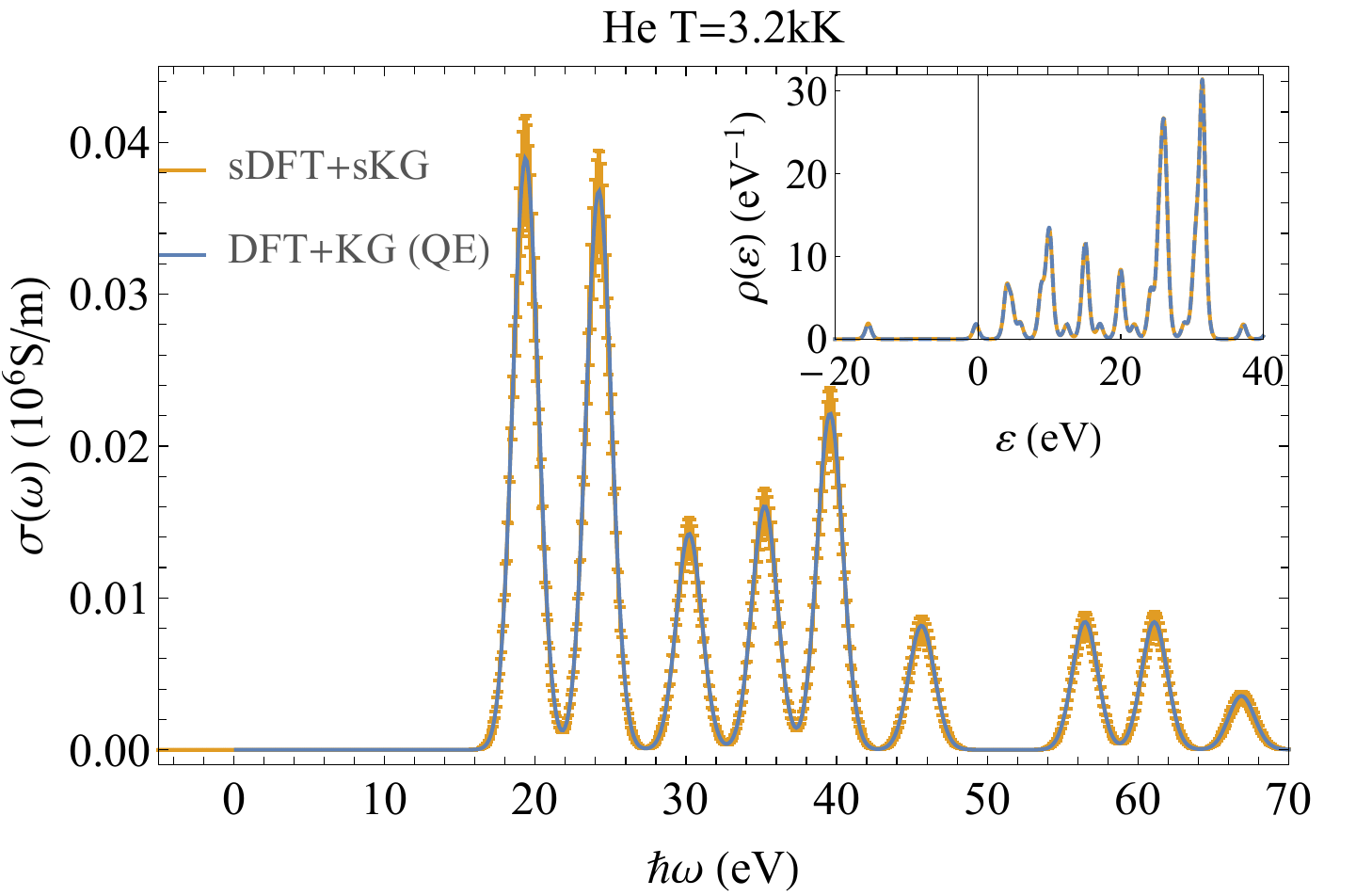}

\includegraphics[width=0.9\columnwidth]{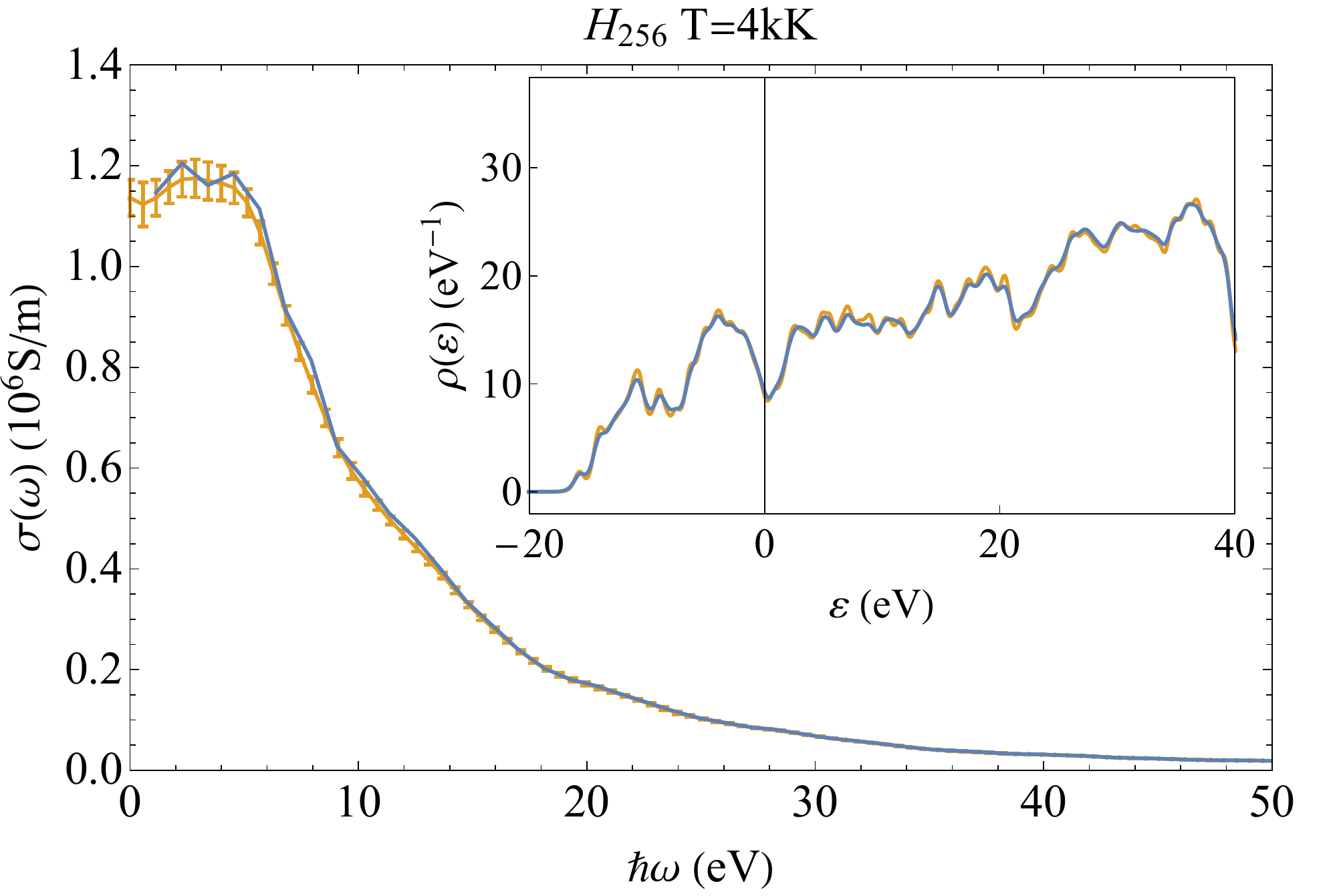}

\caption{\label{fig:he1Compare}\textcolor{black}{{} }The conductivity $\sigma\left(\omega\right)$
and DOS $\rho\left(\varepsilon\right)$ (in the insets) using stochastic
versus deterministic (Quantum Espresso) calculations. We show two
examples, each calculated in a periodic simulation cell of length
$L$ at the $\Gamma$-point: an insulator $\text{He}$ atom at $3200K$
with $L=5.3\angstrom$ (upper panel) and a metallic system, $\text{H}_{256}$
at $4000K$, with $L=8\angstrom$ (lower panel). The deterministic
QE results used $200$ KS eigenstates for the first system and $1700$
for the second. For the stochastic DFT calculation we used 960 stochastic
orbitals for the insulator and 480 orbitals for the conductor. The
conductivity of both systems was calculated using $120$ stochastic
orbitals. A kinetic energy cutoff of $762\;\text{eV}$ for He and
$525$~$\text{eV}$ for $\text{H}_{256}$ was used. Each peak was
Gaussian-broadened, deploying a width parameter equal to $\eta=1.2\text{eV}$
(in sKG this parameter affects Eq. \ref{eq:Cpp}). }
\end{figure}

To validate the method we compare conductivity estimates with that
of the well-established Quantum Espresso (QE) package \citep{Giannozzi2009},
for carrying out both the DFT and the KG (using the KGEC module \citep{calderin2017kubotextendashgreenwood})
calculations. The results for $\text{H}_{256}$ (at $4000K$) and
a single He atom (at $3200K$) are shown in the panels of Fig.~\ref{fig:he1Compare}
and the density of states (DOS) $\rho\left(\varepsilon\right)$, is
shown in the insets. The $\text{H}_{256}$ nuclear configuration was
obtained using an AIMD simulation using VASP. It can be seen that
for both systems the stochastic conductivity spectra and the DOS curves
are in close agreement with the corresponding deterministic estimates
of QE throughout the entire frequency/energy range. 

To further test the method we also looked at $\text{He}_{128}$ systems
in the temperature range $27-57\text{kK}$ and density range $0.71-0.83\frac{\text{g}}{\text{cm}^{3}}$
respectively. To obtain a set of nuclear configurations, an \emph{ab
initio }molecular dynamics (AIMD) trajectory was run using the PBE
\citep{Perdew1996} exchange-correlation (XC) functional, employing
the plane-wave code VASP \citep{Hafner2007,Kresse1996}. Snapshots
of the nuclear configurations were then taken from the equilibrated
part of the simulation, as described in Ref.~\citenum{Preising2018}.
For each snapshot we performed a sDFT calculation using 160 sDFT-os
to obtain the Hamiltonian. The standard deviation was estimated by
using five different sets of 160 sKG-os, each different from the set
used for the Hamiltonian calculation, to avoid bias. For a given nuclear
snapshots, the stochastic calculation produces a conductivity spectrum
with certain stochastic error. The stochastic fluctuations in our
case, turned out to be larger than the fluctuations arising from averaging
over the different nuclear configurations. We therefore present here
results obtained from one snapshot only. To calculate the discretized
momentum-momentum correlation function $C_{PP}\left(\Delta t\times n\right)$
we used $n=600$ time-steps with $\Delta t=0.25\,\hbar E_{h}^{-1}$.
The conductivity spectrum is then obtained from Eq.~\ref{eq:final-sKG-2}
using a Gaussian broadening parameter of $\eta=0.036\text{eV}$.

\begin{figure}[h]
\includegraphics[width=0.9\columnwidth]{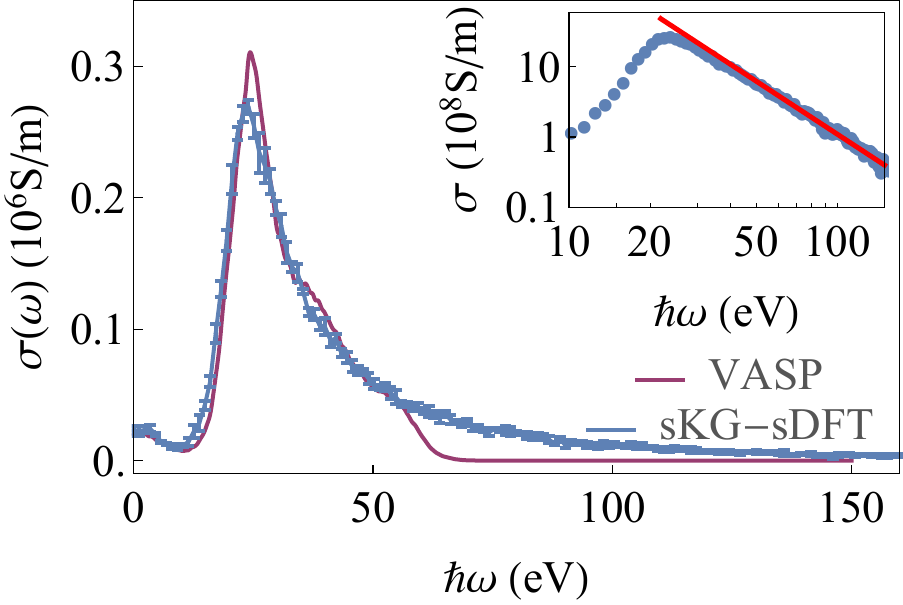}

\caption{\label{fig:BenchFullSpect}The conductivity spectrum for a $\text{He}_{128}$
system at $T=27\text{kK}$ and density of $0.83\frac{\text{g}}{\text{cm}^{3}}$.
The sKG-sDFT LDA conductivity with error bars (with $I_{\sigma}=I_{H}=160$)
is compared to the deterministic results by VASP based on PBE, as
described in Ref. \citenum{Preising2018}. The deterministic calculations
were done with an energy cutoff of $800\text{eV}$ using 570 KS-states.
Inset: The spectrum decay. the red line is proportional to $\omega^{-5/2}$.}
\end{figure}

The full spectrum and the standard deviation involved in the calculation
as described above are shown in Fig.~\ref{fig:BenchFullSpect}. The
advantage of the stochastic method is apparent when looking at frequencies
higher than $\sim70~eV$, where the deterministic calculation of Ref.~\citenum{Preising2018},
gives no contributions above this cut-off energy which has to be introduced
in plane-wave DFT codes like VASP. The deterministic frequency range
could have increased in principle by including more KS states, but
this would require an excessive computational effort. Careful analysis
with respect to the cut-off energy show that equation-of-state data
and the low-frequency conductivity can be converged properly (see
e.g. Refs. \citep{Holst2008,Preising2018}). The sKG calculation on
the other hand samples states from the entire energy spectrum and
therefore exhibits the physically correct asymptotic decay of $\omega^{-5/2}$,
as expected for the free electron gas. The correct high-frequency
asymptotic behavior enables calculation of the Thomas-Reiche-Kuhn
sum-rule \citep{kuhn1925aoeber,reiche1925uberdie} which states that
the total oscillator strength per electron $f_{osc}/N_{e}$, where
\begin{equation}
f_{osc}=\frac{m_{e}\Omega}{\pi e^{2}}\int_{-\infty}^{\infty}\sigma\left(\omega\right)d\omega,\label{eq:f_Osc}
\end{equation}
 and $\sigma\left(\omega\right)$ is the conductivity defined in
Eq.~(\ref{eq:SigmaResponseRelation}), should be equal to 1. The
actual calculated values of $f_{osc}/N_{e}$ are shown in Table~\ref{tab:The-sum-rule}
for three $\text{He}_{128}$ systems (one of which we considered in
Fig.~\ref{fig:BenchFullSpect} and two others, of different temperature
and densities are given for further demonstration) and are indeed
very close to the theoretical value of 1, signifying that the calculations
are converged with respect to the number of states and the total time
of propagation.

At intermediate frequencies, we find (Fig.~\ref{fig:BenchFullSpect})
a close overall agreement between the deterministic and stochastic
estimates of the conductivity spectra, despite the fact that both
methods make use of different XC functionals. The most conspicuous
feature of the spectrum in this range is its peak at $\hbar\omega_{peak}\approx25\text{eV}$,
featuring the maximal deviation between the two spectra which is nonetheless
small, with a 10\% difference in height and 0.3eV difference in the
value of $\hbar\omega_{peak}$.

The DC conductivity for three different systems are displayed in the
third and fourth columns of Table~\ref{tab:The-sum-rule} and the
agreement between the deterministic and stochastic zero frequency
limit is shown.

\begin{center}
\begin{table}[h]
\begin{centering}
\begin{tabular}{ccccc}
\hline 
\multicolumn{2}{c}{System} & \multirow{2}{*}{$f_{osc}/N_{e}$} & \multicolumn{2}{c}{$\sigma_{DC}\left(10^{6}\text{Siemens}/m\right)$}\tabularnewline
\cline{1-2} \cline{2-2} \cline{4-5} \cline{5-5} 
$\rho/(\text{g}/\text{cm}^{3})$ & $T/kK$ &  & VASP/PBE & sKG-sDFT/LDA\tabularnewline
\hline 
$0.71$ & 29 & 1.010 & $0.021\pm0.001$ & $0.026\pm0.002$\tabularnewline
$0.83$ & 27 & 0.998 & $0.018\pm0.001$ & $0.02\pm0.003$\tabularnewline
0.75 & 57 & 1.014 & $0.110\pm0.002$ & $0.10\pm0.02$0\tabularnewline
\hline 
\end{tabular}
\par\end{centering}
\caption{\label{tab:The-sum-rule}The total oscillator strength per electron
$f_{osc}/N_{e}$ (see Eq.\ref{eq:f_Osc})~and the DC conductivity
calculated using VASP based on PBE \citep{Preising2018} and the sDFT-sKG
based on LDA employing $I_{\sigma}=I_{H}=160$ stochastic orbitals.
The statistics for the stochastic calculation is obtained from 5 different
sKG runs and that of the deterministic calculation was taken from
5 points in the proximity of the DC conductivity to evaluate the $\omega\rightarrow0$
limit.}
\end{table}
\par\end{center}

\subsection{\label{subsec:Analysis-of-the}Analysis of the statistical errors}

\begin{figure}
\includegraphics[width=0.9\columnwidth]{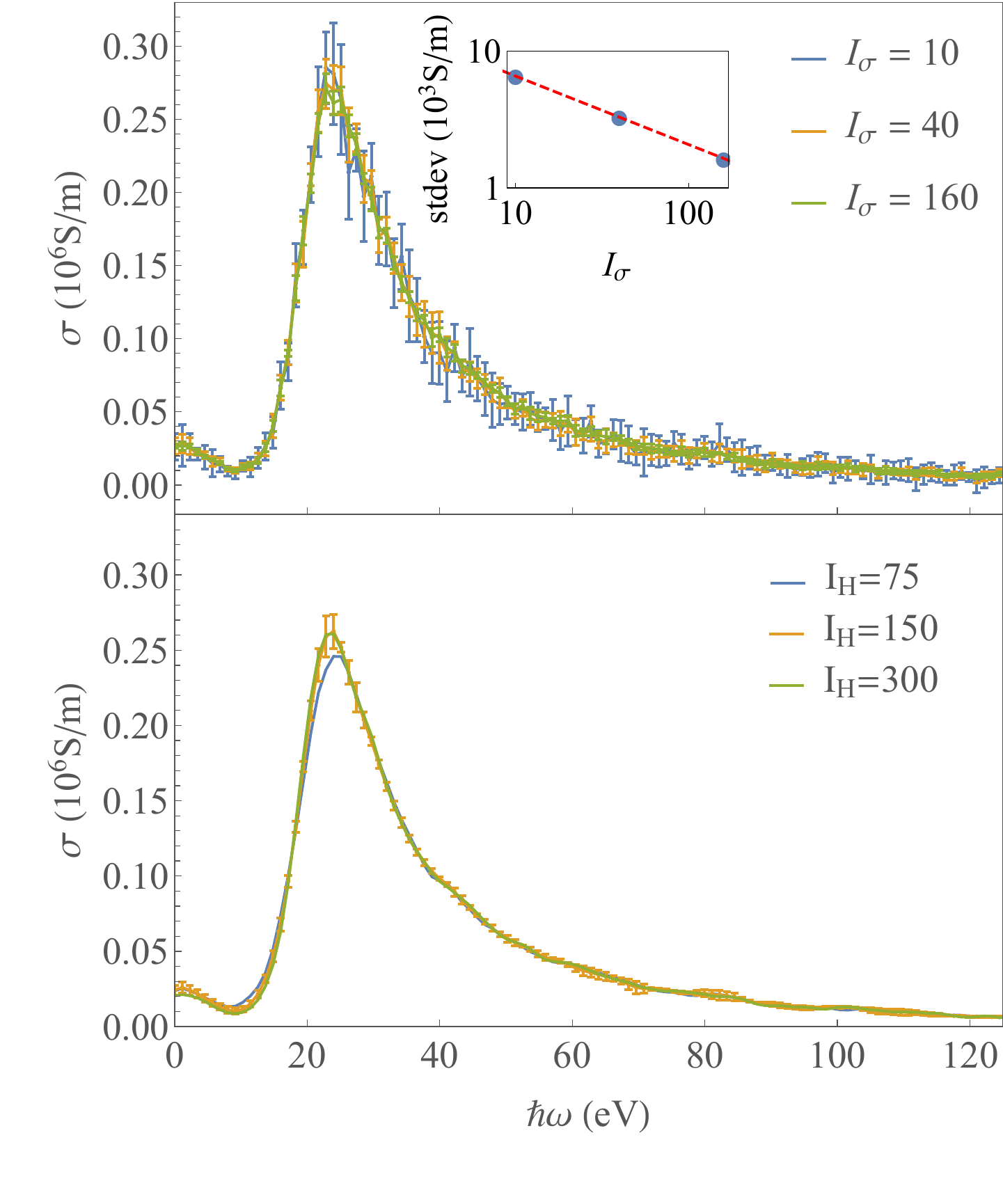}

\caption{\label{fig:he128biasAndfluct}The conductivity spectrum of $\text{He}_{128}$
at 27kK and density of $0.83\text{g/\ensuremath{\text{cm}^{3}}}$.
Top panel: The conductivity based on one sDFT Hamiltonian (using $I_{H}=150$)
calculated with an increasing number $I_{\sigma}$ of sKG-os. Inset:
The standard deviation (stdev) of the conductivity, averaged over
all frequencies as a function of $I_{\sigma}$. The dashed line is
proportional to $I_{\sigma}^{-1/2}$. Bottom panel: The conductivity
based on three sDFT Hamiltonians, each obtained using $I_{H}$ sDFT-os.
In order not to clutter the plot, error bars are given only for the
$I_{H}=150$ sDFT-o's calculation. The sKG calculations were all done
using $I_{\sigma}=150$ sKG-os.}
\end{figure}

There are three sources of statistical errors in the calculation.
The sDFT, that produces the Hamiltonian with which the conductivity
is calculated by Eq.~(\ref{eq:Cpp})-(\ref{eq:sigma(0)}) contributes
two of the errors. One is the fluctuation which is measured by the
standard deviation of the results, and is proportional to $I_{H}^{-1/2}$,
where $I_{H}$ is the number sDFT-os. The second is the bias, related
to the deviance of the average from the exact value, discussed in
previous works \citep{cytter2018stochastic,fabian2018stochastic}
that is proportional to $I_{H}^{-1}$. In addition to the errors in
the sDFT stage, the stochastic evaluation of the momentum-momentum
correlation function also contributes an additional fluctuation. The
effect of the two errors arising from the sDFT calculation on the
conductivity spectrum is displayed at the bottom panel of Fig.~\ref{fig:he128biasAndfluct}.
We show three spectra, each based on a distinct sDFT Hamiltonian,
calculated using different values of $I_{H}$. For the case of $I_{H}=150$
ten different sDFT-o sets were used in order to asses the fluctuation
stemming from the stochastic procedure. For all three conductivity
calculations, we used the same set of $I_{\sigma}=150$ sKG-os, thereby
leading to a similar fluctuation, so that we can focus on the errors
resulting from the sDFT process. The spectra based on $I_{H}=300$
are within the error bars of the $I_{H}=150$ for all frequencies
considered, while the spectrum that is based on $I_{H}=75$ exhibits
a deviation from the other two, especially near the $\omega\sim25eV$
peak. Since the fluctuation is small, we deduce that this difference
can be attributed to the bias component of the statistical error,
and that it is small at $I_{H}=75$ and much smaller than the fluctuation
when $I_{H}\ge150$.

Having discussed the two errors connected with the stochastic nature
of the Hamiltonian, we now address the random fluctuations that arise
from the sKG calculation. For this purpose, we take one of the sDFT
Hamiltonians above (that was calculated using $I_{H}=150$ sDFT-os)
and perform three conductivity spectra calculations on it using different
values $I_{\sigma}$ of sKG-os. The resulting spectra are shown in
the uper panel of Fig. \ref{fig:he128biasAndfluct}. The inset shows
that the standard deviation, averaged over all frequencies, decreases
according to the central limit theorem as expected. Since the sKG-os
are used to directly sample the trace, the statistical error should
be a ``pure'' fluctuation, with no bias. Therefore, while the peak
in this example exhibits a decrease as $I_{\sigma}$ increases, we
attribute that behavior to a fluctuation.

\subsection{\label{subsec:Algorithmic-implementation-and}Algorithmic implementation
and scaling of the algorithm}

The computational time of the sKG algorithm, as described in subsection
\ref{subsec:Stochastic-response-function} and Eq.~(\ref{eq:symResponse})
is determined mainly by application of $f_{FD}\left(\hat{h}\right)$,
$\sqrt{f_{FD}\left(\hat{h}\right)}$ and the time evolution operator
$e^{-i\hat{h}\Delta t}$, all functions of the Hamiltonian $\hat{h}$
on given wave-functions. Each of these Hamiltonian functions can be
applied by using Chebyshev expansions \citep{Kosloff1988,Goedecker1994,Huang1995,Baer1997a},
where the Hamiltonian is applied to the wave function repeatedly $N_{C}$
times. The length of the expansion $N_{C}$ is proportional to $\Delta E=E_{max}-E_{min}$
where $E_{max}$ and $E_{min}$ are upper and lower bounds on the
maximal and minimal eigenvalues of $\hat{h}$ respectively. For the
Fermi-Dirac functions $f_{FD}\left(\hat{h}\right)$ and $\sqrt{f_{FD}\left(\hat{h}\right)}$
the Chebyshev expansion length $N_{C}$ is proportional to $\beta\Delta E$.

\begin{figure}
\includegraphics[width=0.9\columnwidth]{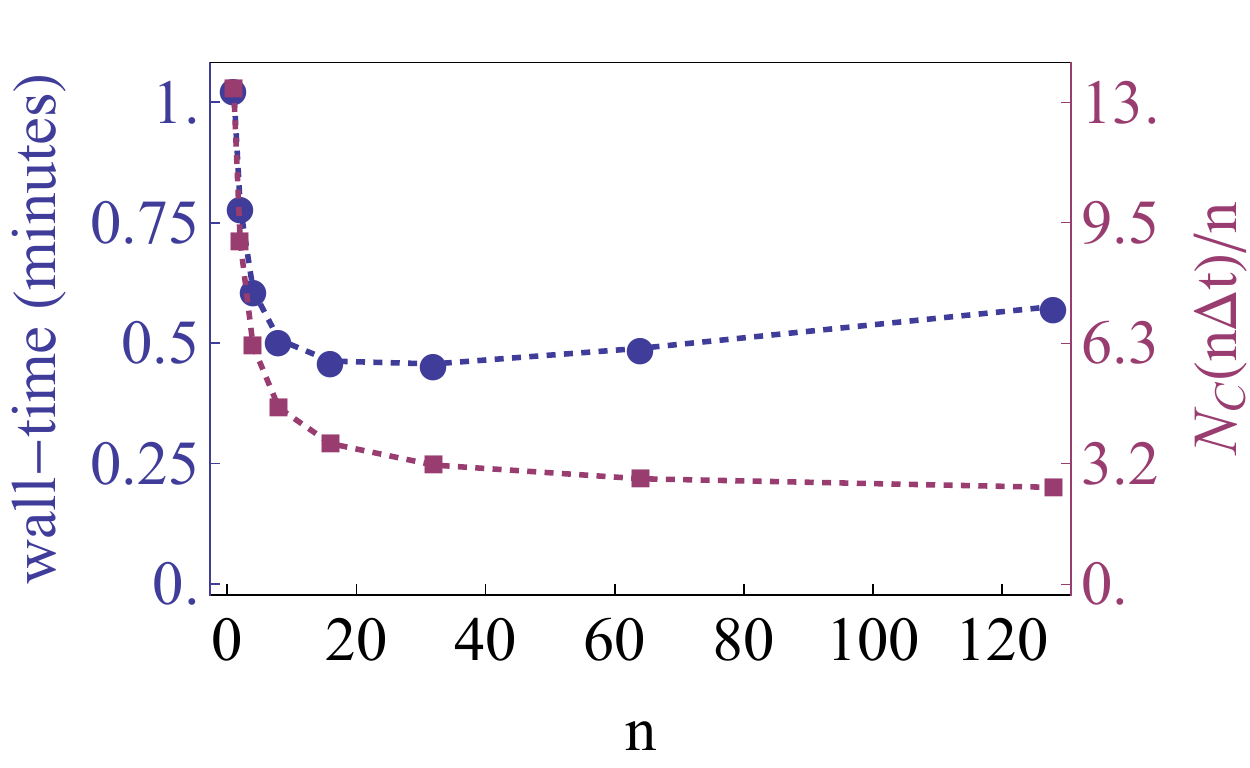}

\caption{\label{fig:timeCheby}The CPU time (blue circled markers) and $N_{C}\left(n\Delta t\right)/n$
(purple square markers) as a function of the number $n$ of time step
propagation operators used in the calculation. The calculation was
done for $\text{He}_{128}$ using $N_{g}=60^{3}$ grid points, at
$57\text{kK}$ on 8 processors where $N_{ts}=128$ and $\Delta t=0.25au$,
for one dipole direction.}
\end{figure}

Propagating the wave function $\varphi$ to different times can be
performed with several Chebyshev expansions:
\begin{equation}
\varphi_{n}=e^{-i\hat{h}\left(n\Delta t\right)}\varphi=\sum_{m=0}^{N_{C}\left(n\Delta t\right)}a_{m}\left(n\Delta t\right)\phi_{m},\label{eq:Chebysh}
\end{equation}
where $\phi_{m}$ are the Chebyshev recursion wave functions \footnote{The Chebyshev recursion is $\phi_{m+1}=2\hat{h}_{N}\phi_{m}-\phi_{m-1}$,
with $\phi_{0}=\varphi$ and $\phi_{1}=\hat{h}_{N}\phi_{0}$, where
$\hat{h}_{N}=\frac{\hat{h}-\bar{E}}{\Delta E}$, $\bar{E}=\frac{1}{2}\left(E_{max}+E_{min}\right)$
and $\Delta E=\frac{1}{2}\left(E_{max}-E_{min}\right)$.}. Note that $\varphi_{n}$ for the different values of $n$ are different
linear combinations of the same recursion wave functions $\phi_{m}$,
but summed with different expansion coefficient $a_{m}\left(n\Delta t\right)$.
We can therefore generate one set of $\phi_{1},\dots,\phi_{N_{C}}$
for generating the required set of $\varphi_{n}'s$. The Chebyshev
expansion length $N_{C}$ is determined as the smallest integer for
which $\left|a_{m}\left(n\Delta t\right)\right|<10^{-9}$ for all
$m\ge N_{C}$. Clearly, $N_{C}$ depends on $n\Delta t$, hence the
notation $N_{C}\left(n\Delta t\right)$. The expansions in Eq.~(\ref{eq:Chebysh})
are highly beneficial since most of the computational effort goes
to applying the Hamiltonian on the different wave-functions, that
is, calculating the set of $\phi's$. Thus, to find the optimized
number of simultaneously calculated time-steps $n$, in Fig.~\ref{fig:timeCheby}
we look at the number of Chebyshev terms required \emph{per time step},
$N_{C}\left(n\Delta t\right)/n$, along side the wall time for every
choice of $n$. It can be seen that $N_{C}\left(n\Delta t\right)/n$
is highest at $n=1$ and as $n$ increases, its value drops steeply
towards an asymptotic plateau value smaller by a factor of $\backsim4$.
It is seen that using this approach CPU times indeed decrease but
due to an additional overhead of the calculation only a factor of
2 is obtained.

\begin{figure}
\includegraphics[width=0.9\columnwidth]{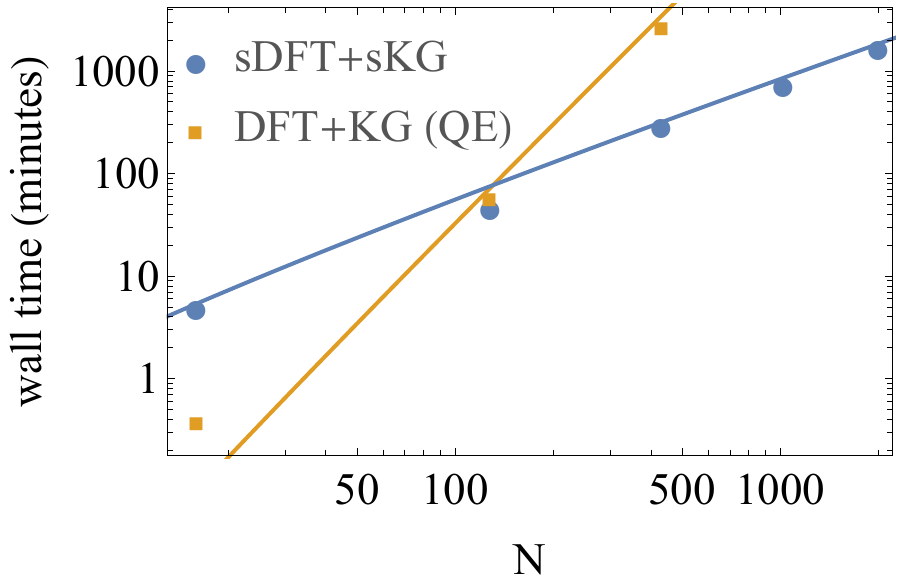}

\caption{\label{fig:cpuFig}Wall time for stochastic DFT+KG calculations with
600 time steps with $\Delta t=0.25\,\hbar E_{h}^{-1}$ as well as
a deterministic calculation done using Quantum Espresso (QE), as a
function of the number of $\text{He}$ atoms, at $0.75\frac{\text{g}}{\text{cm}^{3}}$
and 57kK. The orange curve is proportional to $N^{3}$, and the blue
curve is linear with $N$, the number of atoms.}
\end{figure}

The computational effort for the sKG procedure has a near-linear scaling
with system size $N$ as does the sDFT , and this is due to the following
two reasons: 1) the Hamiltonian $\hat{h}$ action on a wave function
involves a $O\left(N_{g}\ln N_{g}\right)$ numerical complexity (this
is the operation count of the fast Fourier transform involved in the
kinetic energy operation), where $N_{g}\propto N$ is the number of
grid-points; and 2) The number of such Hamiltonian operations is $N_{C}\times I_{\sigma}$,
where $N_{C}$ (the Chebyshev expansion length) and $I_{\sigma}$
(the number of sKG-os) are both system-size independent. For the same
reasons t the sDFT calculation also scales linearly with $N$ (as
shown before in Ref. \citenum{Baer2013}). Furthermore, the computational
effort in sDFT decreases as the temperature increases in proportion
to $1/T$ \citep{cytter2018stochastic}, due to the fact that the
FD Chebyshev expansion length $N_{C}$ is proportional to $\Delta E/k_{B}T$
\citep{baer1997sparsity} and the energy range $\Delta E$ is system-size
independent. The $O\left(N/T\right)$ scaling with system size and
temperature we report here should be compared to the $O\left(N^{3}T^{3}\right)$
scaling of the deterministic calculation based on Eq.~(\ref{eq:conduct_deterministic}),
which requires calculation of all the occupied (and many unoccupied)
states, the number of which is proportional to $T^{3}$ (based on
the electron gas density of states). The system-size scaling can be
seen in actual calculation, as shown in Fig.$\ $\ref{fig:cpuFig},
where the wall-time for the DFT+KG calculation, is shown as a function
of the number of atoms in the system (keeping the density and temperature
fixed as the number of atoms increases) for deterministic (using QE)
and sDFT+sKG. For small system sizes the QE calculation is considerably
faster than the stochastic approach. However, as the system size increases,
due to liner scaling, the stochastic approach becomes competitive.
At $N=128$ we find a crossover and already for $N=432$ the stochastic
calculation is 10 times faster than the deterministic one. 

\section{\label{sec:Mixed-systems}Mixed He/H WDM systems}

\begin{figure}
\includegraphics[width=1\columnwidth]{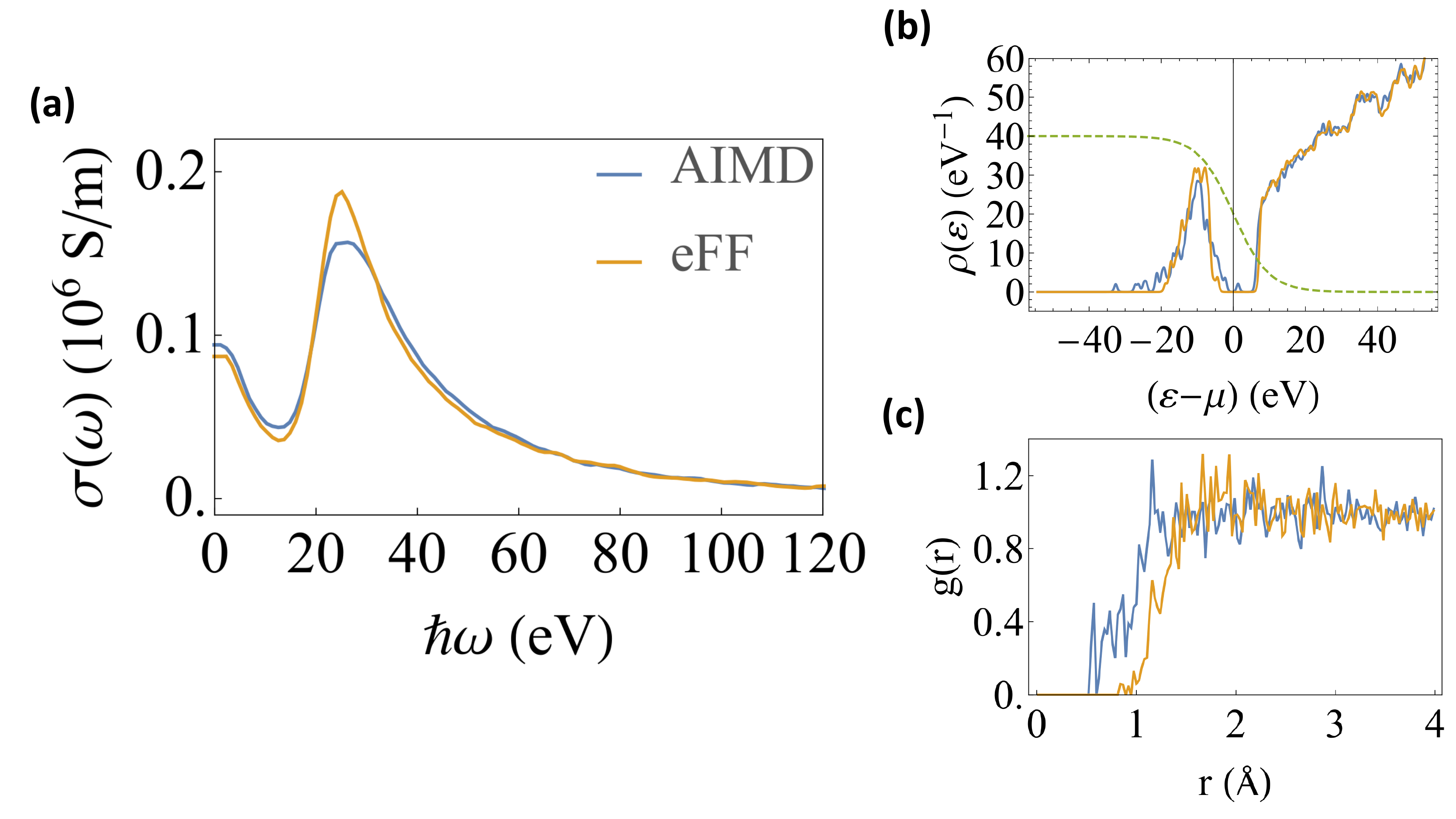}

\caption{\textcolor{blue}{\label{fig:lammpsCompare}} Comparison of the calculated
conductivity $\sigma\left(\omega\right)$ (a), the DOS $\rho\left(\varepsilon\right)$
(b) and the radial pair correlation $g\left(r\right)$ (c) for $\text{He}_{128}$
at 57kK and density of $0.75\frac{\text{g}}{\text{cm}^{3}}$ based
on configurations generated by AIMD \citep{Preising2018} vs. eFF
dynamics. The dashed line in panel (b) shows the Fermi-Dirac level
occupation, transiting from a value of 1 at low energies $\varepsilon$
to $0$ at high energies .}
\end{figure}

As an application of the method, we study the conductivity and DOS
of various systems with different hydrogen-helium mixtures at temperature
of $57\text{kK}$ and constant volume. For each system, we obtained
a set of thermally-distributed nuclear configurations using the electron
force-field (eFF) \citep{JaramilloBotero2010} dynamics as implemented
in LAMMPS \citep{plimpton1995fastparallel}, which has been shown
has been shown to give a good description of the pair correlation
and equations of state of first-row materials under extreme conditions
\citep{kim2011hightemperature,su2007excited}. For $\text{He}_{128}$
at $57\text{kK}$ we generate a set of Boltzmann-distributed configurations
using both an empirical force field and an ab-inito approach taken
from Ref.~\citenum{Preising2018}. The configurations where then
used to average the results over the thermal fluctuations of the nuclei.
In Fig.~\ref{fig:lammpsCompare} we compare the two sets of results
and show that while giving two visibly different spectra they share
similar trends with peaks/troughs located at nearly identical frequencies.
Comparing the two DOSs we see small differences in the occupied state
energies while being nearly identical at the unoccupied state energies.
Comparing the correlation functions $g\left(r\right)$, we find that
AIMD gives significant weight to He pairs approaching as close as $0.5\angstrom$ while the eFF does not. Both functions show a peak
at $1.1\angstrom$, but it is more significant in AIMD. It
is perhaps surprising that despite the rather large differences in
the pair correlation between the two methods, the electronic properties,
as mentioned above, are not very different.

We characterize the mixture by the hydrogen fraction in the system

\begin{equation}
X_{\text{H}}=\frac{N_{\text{H}}}{N_{\text{H}}+N_{\text{He}}}\ ,\label{eq:chiH-1}
\end{equation}
where $N_{\text{H}}$ and $N_{\text{He}}$ are the number of hydrogen
and helium atoms, respectively. For practical purposes, this ratio
is achieved by holding the total number of atoms $N_{\text{H}}+N_{\text{He}}$
in the simulation cell constant and equal to $1024$.

We ran five molecular dynamics trajectories at fixed volume (minimum
image periodic boundary conditions for $L=39.4\,a_{0}$) and temperature
($T=57\text{kK}$) with interactions between He and H described by
the eFF force-field with a cutoff of 6.45$a_{0}$. Each trajectory
started with the same ordered configuration, and a different velocity
allocation, equilibrated, and then ran for a total of 3ps with time
step of $10^{-3}\text{fs}$, needed due to consideration of both electronic
and nuclear time scales. The duration of the trajectories corresponded
to the correlation time of 3ps estimated using the same data. The
final nuclear configuration for each trajectory represented a set
of five uncorrelated H-He mixtures. For each structure, a sDFT calculation
determined the Hamiltonian $\hat{h}$ which was used for the sKG calculation
of the conductivity spectrum. For both sDFT and sKG an identical simulation
box and grid of $N_{g}=120^{3}$ points was used which correspond
to a grid spacing of $\delta x=0.33\;a_{0}$. The sDFT calculation
was based on $I_{H}=120$ sDFT-o's and the sKG calculation used a
distinct set of $I_{\sigma}=120$ sKG-o's.

\begin{figure}
\includegraphics[width=0.9\columnwidth]{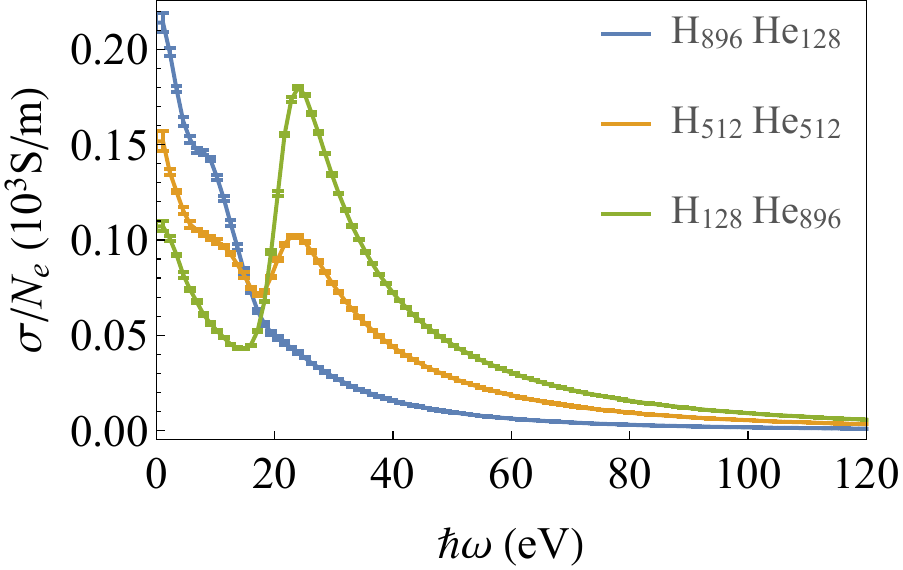}

\includegraphics[width=0.9\columnwidth]{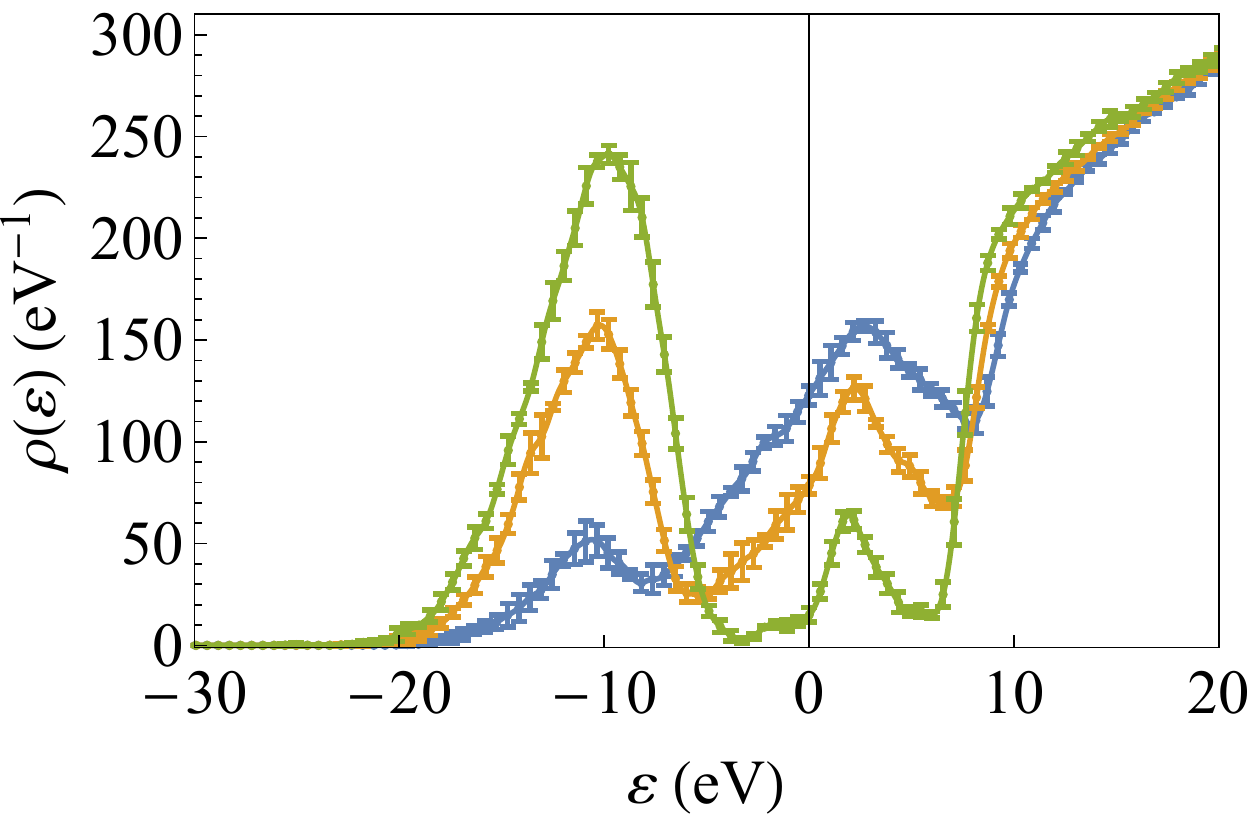}

\caption{\label{fig:dosAndConductDiffHP}The Conductivity (upper panel) and
the DOS (lower panel) of different systems containing 1024 atoms with
different hydrogen percentages at $T=57\text{kK}$ and an average
atomic volume of $60\,a_{0}^{3}$ per atom. The conductivity was normalized
according to the number of electrons in the system. The DOS is shifted
so that the chemical potential is zero.}
\end{figure}

In Fig.~\ref{fig:dosAndConductDiffHP} the conductivity spectra and
the density of states (DOS) for three different mixtures is displayed.
These two characteristics are closely related and will therefore be
discussed together. The statistical fluctuation in the DOS (lower
panel), denoted as error bars, was determined by running sDFT calculations
on the five distinct configuration snapshots as described above, each
using a different set of sDFT-o's. These five Hamiltonians were then
used for evaluating the conductivity (upper panel) employing a different
set of sKG-o's to avoid additional bias. The resulting five conductance
spectra and DOS were used for estimating the thermally-averaged curves
and their associated statistical errors. It is seen in the upper panel
of Fig.~\ref{fig:dosAndConductDiffHP}, that the statistical fluctuations
are small compared to the difference between the curves and they do
not seem to increase as a function of the hydrogen atomic fraction
$X_{H}$ and therefore, only one snapshot was used in all other calculations.

When a relatively small fraction of hydrogen atoms is present in the
system, it gives rise to a small peak at 3eV inside the Helium energy
gap in the DOS (see the lower panel of Fig.~\ref{fig:dosAndConductDiffHP}).
As the hydrogen concentration increases the He gap fills with states
until it is no longer visible and at the same time the DOS of the
valence band (seen in the figure at around $-10\text{eV}$), decreases
steadily. Both effects show a gradual transition to metallization
as the hydrogen ratio grows. At high energies the DOS of all mixtures
converges to the free electron limit.

\begin{figure}
\includegraphics[width=0.9\columnwidth]{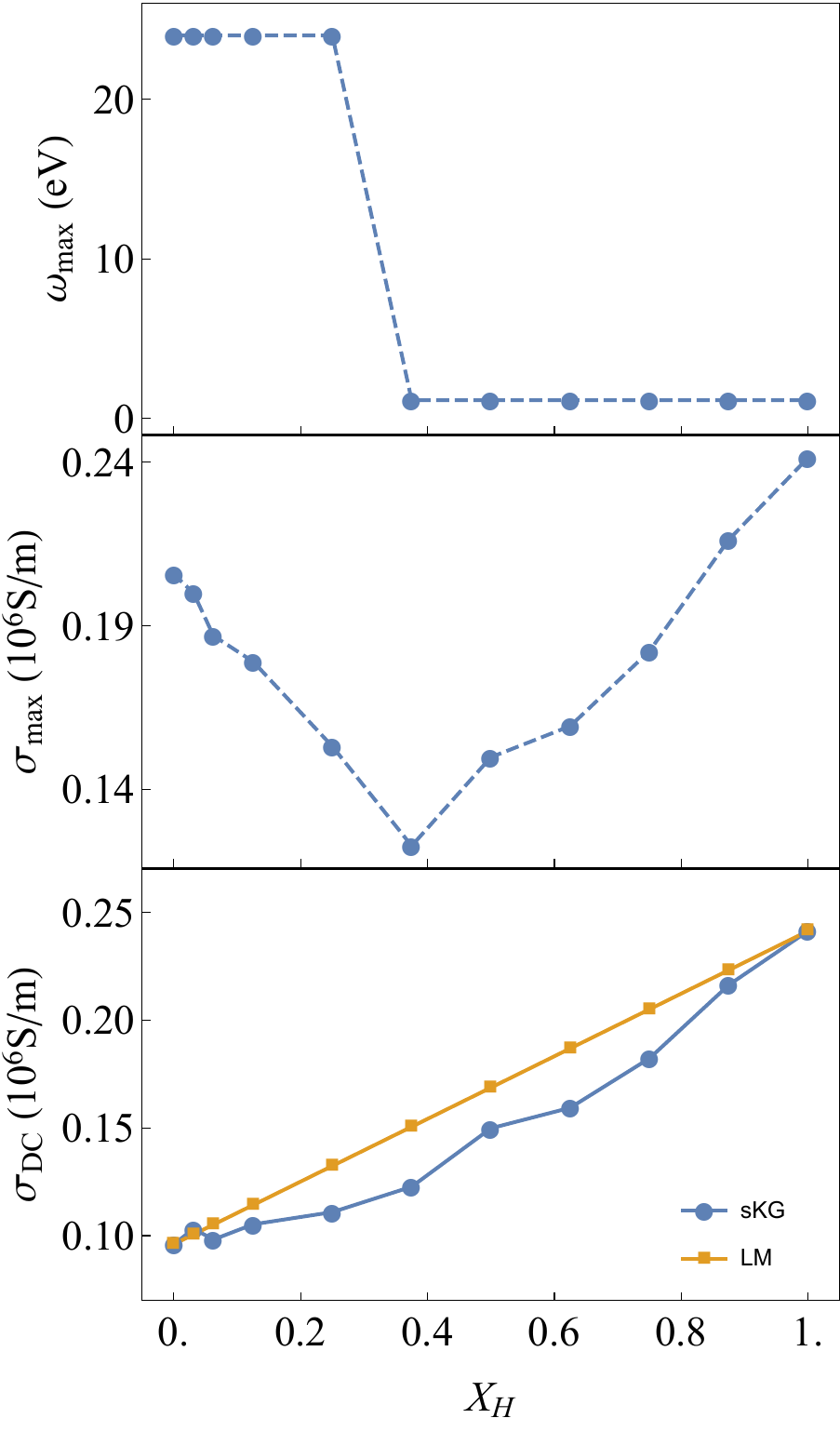}

\caption{\label{fig:maxSigAndDC}The maximal conductivity frequency $\omega_{max}$
(top panel), maximal conductivity $\sigma_{max}$ (middle panel) and
the DC conductivity of the actual (round blue markers) and the linear-mixing
model $\sigma_{\text{LM}}$ (square orange markers), as a function
of the hydrogen ratio $X_{H}$ in He-H mixtures.}
\end{figure}

The sKG conductivity follows the changes seen in the DOS. Consider
first the DC conductivity, shown in the lower panel of Fig.~\ref{fig:maxSigAndDC},
which remains relatively constant as the hydrogen fraction grows until
$X_{H}^{crit}\sim0.3$. Beyond this value of the hydrogen fraction
the DC conductivity increases near-linearly with $\chi_{\text{H}}$
as a result of the energy gap filling in the DOS, allowing more transitions
at low energies. Due to the finite temperature and therefore partial
occupation there exists zero frequency transitions even at helium
dominated systems causing the DC conductivity to change only by a
factor of 2.5 when moving from pure helium to pure hydrogen systems.
The peak in the He dominated spectrum, as seen in the upper panel
of Fig.~\ref{fig:dosAndConductDiffHP}, appears at around $25\text{eV}$
and corresponds to the transition from the highest density in the
occupied band to the non-occupied band threshold levels (as seen in
the DOS at $10\text{eV}$). Furthermore, at higher $\text{He}$ concentrations
due to the energy gap, transitions in 15eV become less probable, resulting
in a local minimum in the conductivity at this frequency.

Next, we consider the frequency $\omega_{max}$ for which the conductivity
is maximal, plotted as a function of $X_{H}$ in the top panel of
Fig.~\ref{fig:maxSigAndDC}. This frequency displays an abrupt shift
of $\omega_{max}$ from $\sim25\text{eV}$ to $0$ (DC) as $X_{H}$
passes through the critical value of $X_{H}^{crit}\sim0.3$. This
critical value, indicates an abrupt nonmetal-to-metal transition in
the H-He system as reported in Ref.~\citenum{lorenzen2011metallization}
for considerably lower temperatures. This critical hydrogen concentration
is well withing the range of the Mott criterion for metallization
in pure hydrogen, as seen in Ref.~\citenum{lorenzen2009demixing}
that shows it occurs at $n_{H}^{1/3}a_{0}\approx0.25$ for temperatures
up to 15kK. In the present system, we find the metallization density
at $n_{H}^{1/3}a_{0}\approx0.18$, which seems reasonable considering
the fact that we're looking at a substantially higher temperature
in which thermal effects promote the conductivity onset.

The finite $\omega_{max}$ is a results of the energy gap in what
is generally an insulating system (He dominated) and the zero $\omega_{max}$
signifies its disappearance, allowing many of the energy transitions
to occur at infinitesimal energy values. In the middle panel of Fig.
\ref{fig:maxSigAndDC} the transition through $X_{H}^{crit}$ is seen
as a qualitative change in the behavior of the maximal conductivity
$\sigma_{max}$, which initially decreases as $X_{H}$ approaches
$X_{H}^{crit}$, and then increases as $X_{H}$ grows further.

Finally, we compare the spectra in the different concentrations to
a model of linear averaging of pure helium and pure hydrogen spectra,
defined as
\begin{equation}
\sigma_{\text{LM}}\left(X_{\text{H}};\omega\right)=X_{\text{H}}\sigma\left(1;\omega\right)+\left(1-X_{\text{H}}\right)\sigma\left(0;\omega\right)\;.\label{eq:linearMixCond}
\end{equation}
It can be seen in the lower panel of Fig.~\ref{fig:maxSigAndDC}
that the DC conductivity $\sigma_{LM}\left(X_{\text{H}};0\right)$
based on the linear averaging model is typically greater than the
corresponding value calculated using sKG . This is due to the fact
that in the actual system the environment each atom experiences includes,
on the average, a mixture of He and H atoms while in the linear averaging
model each atom is surrounded by atoms of its own kind.

\section{Summary and Conclusions}

In this work we presented a stochastic approach, sKG, to calculate
the conductivity using Kubo-Greenwood formalism on top of a sDFT calculation.
We showed that sKG conductivity can approach the values of the deterministic
KS conductivity determined by the KG method when the number of sDFT-o's
and sKG-o's are increased systematically. Moreover, the uniform sampling
of all states of the system by sKG allows it to describe equally well
the low-, mid- and high-end ranges of the spectrum, while the deterministic
method is limited to lower energies due to memory and CPU constraints.
The computational effort of the method scales linearly with system
size and inversely proportional to the temperature similar to the
sDFT calculations \citep{cytter2018stochastic} while the deterministic
approach has cubic scaling both in system size and temperature.

As an application of the method, we studied the conductivity and DOS
for mixed hydrogen and helium systems at a constant volume and temperature
($T=57\text{kK}$) ensemble. We found that the system displays two
conductivity phases, where a transition from insulator to metal occurs
at hydrogen atomic fraction of $X_{H}\approx0.3$.

The method enlarges the scope of sDFT to study properties of warm
dense matter for very large systems at high temperatures. This could
be significant when large inhomogeneous systems are studied or in
systems where the mixing occurs on the nanoscale.
\begin{acknowledgments}
RB thanks the  US-Israel Binational Science Foundation grant number BSF-2018368. RR thanks
the DFG for support within the FOR 2440. DN and ER are grateful for
support by the Center for Computational Study of Excited State Phenomena
in Energy Materials (C2SEPEM) at the Lawrence Berkeley National Laboratory,
which is funded by the U.S. Department of Energy, Office of Science,
Basic energy Sciences, Materials Sciences and Engineering Division
under contract No. DEAC02-05CH11231 as part of the Computational Materials
Sciences Program.
\end{acknowledgments}

\bibliographystyle{apsrev4-2}

\begin{thebibliography}{53}%
\makeatletter
\providecommand \@ifxundefined [1]{%
 \@ifx{#1\undefined}
}%
\providecommand \@ifnum [1]{%
 \ifnum #1\expandafter \@firstoftwo
 \else \expandafter \@secondoftwo
 \fi
}%
\providecommand \@ifx [1]{%
 \ifx #1\expandafter \@firstoftwo
 \else \expandafter \@secondoftwo
 \fi
}%
\providecommand \natexlab [1]{#1}%
\providecommand \enquote  [1]{``#1''}%
\providecommand \bibnamefont  [1]{#1}%
\providecommand \bibfnamefont [1]{#1}%
\providecommand \citenamefont [1]{#1}%
\providecommand \href@noop [0]{\@secondoftwo}%
\providecommand \href [0]{\begingroup \@sanitize@url \@href}%
\providecommand \@href[1]{\@@startlink{#1}\@@href}%
\providecommand \@@href[1]{\endgroup#1\@@endlink}%
\providecommand \@sanitize@url [0]{\catcode `\\12\catcode `\$12\catcode
  `\&12\catcode `\#12\catcode `\^12\catcode `\_12\catcode `\%12\relax}%
\providecommand \@@startlink[1]{}%
\providecommand \@@endlink[0]{}%
\providecommand \url  [0]{\begingroup\@sanitize@url \@url }%
\providecommand \@url [1]{\endgroup\@href {#1}{\urlprefix }}%
\providecommand \urlprefix  [0]{URL }%
\providecommand \Eprint [0]{\href }%
\providecommand \doibase [0]{https://doi.org/}%
\providecommand \selectlanguage [0]{\@gobble}%
\providecommand \bibinfo  [0]{\@secondoftwo}%
\providecommand \bibfield  [0]{\@secondoftwo}%
\providecommand \translation [1]{[#1]}%
\providecommand \BibitemOpen [0]{}%
\providecommand \bibitemStop [0]{}%
\providecommand \bibitemNoStop [0]{.\EOS\space}%
\providecommand \EOS [0]{\spacefactor3000\relax}%
\providecommand \BibitemShut  [1]{\csname bibitem#1\endcsname}%
\let\auto@bib@innerbib\@empty
\bibitem [{\citenamefont {Militzer}(2013)}]{militzer2013equation}%
  \BibitemOpen
  \bibfield  {author} {\bibinfo {author} {\bibfnamefont {B.}~\bibnamefont
  {Militzer}},\ }\href {https://doi.org/10.1103/PhysRevB.87.014202} {\bibfield
  {journal} {\bibinfo  {journal} {Phys. Rev. B}\ }\textbf {\bibinfo {volume}
  {87}},\ \bibinfo {pages} {014202} (\bibinfo {year} {2013})}\BibitemShut
  {NoStop}%
\bibitem [{\citenamefont {Lorenzen}\ \emph {et~al.}(2011)\citenamefont
  {Lorenzen}, \citenamefont {Holst},\ and\ \citenamefont
  {Redmer}}]{lorenzen2011metallization}%
  \BibitemOpen
  \bibfield  {author} {\bibinfo {author} {\bibfnamefont {W.}~\bibnamefont
  {Lorenzen}}, \bibinfo {author} {\bibfnamefont {B.}~\bibnamefont {Holst}},\
  and\ \bibinfo {author} {\bibfnamefont {R.}~\bibnamefont {Redmer}},\ }\href
  {https://doi.org/10.1103/PhysRevB.84.235109} {\bibfield  {journal} {\bibinfo
  {journal} {Phys. Rev. B}\ }\textbf {\bibinfo {volume} {84}},\ \bibinfo
  {pages} {235109} (\bibinfo {year} {2011})}\BibitemShut {NoStop}%
\bibitem [{\citenamefont {Sch{\"o}ttler}\ and\ \citenamefont
  {Redmer}(2018)}]{schottler2018abinitio}%
  \BibitemOpen
  \bibfield  {author} {\bibinfo {author} {\bibfnamefont {M.}~\bibnamefont
  {Sch{\"o}ttler}}\ and\ \bibinfo {author} {\bibfnamefont {R.}~\bibnamefont
  {Redmer}},\ }\href {https://doi.org/10.1103/PhysRevLett.120.115703}
  {\bibfield  {journal} {\bibinfo  {journal} {Phys. Rev. Lett.}\ }\textbf
  {\bibinfo {volume} {120}},\ \bibinfo {pages} {115703} (\bibinfo {year}
  {2018})}\BibitemShut {NoStop}%
\bibitem [{\citenamefont {Stevenson}(1975)}]{Stevenson1975}%
  \BibitemOpen
  \bibfield  {author} {\bibinfo {author} {\bibfnamefont {D.~J.}\ \bibnamefont
  {Stevenson}},\ }\href {https://doi.org/10.1103/PhysRevB.12.3999} {\bibfield
  {journal} {\bibinfo  {journal} {Phys. Rev. B}\ }\textbf {\bibinfo {volume}
  {12}},\ \bibinfo {pages} {3999} (\bibinfo {year} {1975})}\BibitemShut
  {NoStop}%
\bibitem [{\citenamefont {Nettelmann}\ \emph {et~al.}(2008)\citenamefont
  {Nettelmann}, \citenamefont {Holst}, \citenamefont {Kietzmann}, \citenamefont
  {French}, \citenamefont {Redmer},\ and\ \citenamefont
  {Blaschke}}]{nettelmann2008abinitio}%
  \BibitemOpen
  \bibfield  {author} {\bibinfo {author} {\bibfnamefont {N.}~\bibnamefont
  {Nettelmann}}, \bibinfo {author} {\bibfnamefont {B.}~\bibnamefont {Holst}},
  \bibinfo {author} {\bibfnamefont {A.}~\bibnamefont {Kietzmann}}, \bibinfo
  {author} {\bibfnamefont {M.}~\bibnamefont {French}}, \bibinfo {author}
  {\bibfnamefont {R.}~\bibnamefont {Redmer}},\ and\ \bibinfo {author}
  {\bibfnamefont {D.}~\bibnamefont {Blaschke}},\ }\href
  {https://doi.org/10.1086/589806} {\bibfield  {journal} {\bibinfo  {journal}
  {ApJ}\ }\textbf {\bibinfo {volume} {683}},\ \bibinfo {pages} {1217} (\bibinfo
  {year} {2008})}\BibitemShut {NoStop}%
\bibitem [{\citenamefont {Guillot}(1999)}]{Guillot1999}%
  \BibitemOpen
  \bibfield  {author} {\bibinfo {author} {\bibfnamefont {T.}~\bibnamefont
  {Guillot}},\ }\href {https://doi.org/10.1126/science.286.5437.72} {\bibfield
  {journal} {\bibinfo  {journal} {Science}\ }\textbf {\bibinfo {volume}
  {286}},\ \bibinfo {pages} {72} (\bibinfo {year} {1999})}\BibitemShut
  {NoStop}%
\bibitem [{\citenamefont {Silvestrelli}\ \emph {et~al.}(1996)\citenamefont
  {Silvestrelli}, \citenamefont {Alavi}, \citenamefont {Parrinello},\ and\
  \citenamefont {Frenkel}}]{Silvestrelli1996}%
  \BibitemOpen
  \bibfield  {author} {\bibinfo {author} {\bibfnamefont {P.~L.}\ \bibnamefont
  {Silvestrelli}}, \bibinfo {author} {\bibfnamefont {A.}~\bibnamefont {Alavi}},
  \bibinfo {author} {\bibfnamefont {M.}~\bibnamefont {Parrinello}},\ and\
  \bibinfo {author} {\bibfnamefont {D.}~\bibnamefont {Frenkel}},\ }\href@noop
  {} {\bibfield  {journal} {\bibinfo  {journal} {Phys. Rev. Lett.}\ }\textbf
  {\bibinfo {volume} {77}},\ \bibinfo {pages} {3149} (\bibinfo {year}
  {1996})}\BibitemShut {NoStop}%
\bibitem [{\citenamefont {Mattsson}\ and\ \citenamefont
  {Wahnstrom}(1997)}]{Mattsson1997}%
  \BibitemOpen
  \bibfield  {author} {\bibinfo {author} {\bibfnamefont {T.~R.}\ \bibnamefont
  {Mattsson}}\ and\ \bibinfo {author} {\bibfnamefont {G.}~\bibnamefont
  {Wahnstrom}},\ }\href@noop {} {\bibfield  {journal} {\bibinfo  {journal}
  {Phys. Rev. B}\ }\textbf {\bibinfo {volume} {56}},\ \bibinfo {pages} {14944}
  (\bibinfo {year} {1997})}\BibitemShut {NoStop}%
\bibitem [{\citenamefont {Pozzo}\ \emph {et~al.}(2012)\citenamefont {Pozzo},
  \citenamefont {Davies}, \citenamefont {Gubbins},\ and\ \citenamefont
  {Alf{\`e}}}]{pozzo2012thermal}%
  \BibitemOpen
  \bibfield  {author} {\bibinfo {author} {\bibfnamefont {M.}~\bibnamefont
  {Pozzo}}, \bibinfo {author} {\bibfnamefont {C.}~\bibnamefont {Davies}},
  \bibinfo {author} {\bibfnamefont {D.}~\bibnamefont {Gubbins}},\ and\ \bibinfo
  {author} {\bibfnamefont {D.}~\bibnamefont {Alf{\`e}}},\ }\href
  {https://doi.org/10.1038/nature11031} {\bibfield  {journal} {\bibinfo
  {journal} {Nature}\ }\textbf {\bibinfo {volume} {485}},\ \bibinfo {pages}
  {355} (\bibinfo {year} {2012})}\BibitemShut {NoStop}%
\bibitem [{\citenamefont {Witte}\ \emph {et~al.}(2018)\citenamefont {Witte},
  \citenamefont {Sperling}, \citenamefont {French}, \citenamefont {Recoules},
  \citenamefont {Glenzer},\ and\ \citenamefont
  {Redmer}}]{witte2018observations}%
  \BibitemOpen
  \bibfield  {author} {\bibinfo {author} {\bibfnamefont {B.~B.~L.}\
  \bibnamefont {Witte}}, \bibinfo {author} {\bibfnamefont {P.}~\bibnamefont
  {Sperling}}, \bibinfo {author} {\bibfnamefont {M.}~\bibnamefont {French}},
  \bibinfo {author} {\bibfnamefont {V.}~\bibnamefont {Recoules}}, \bibinfo
  {author} {\bibfnamefont {S.~H.}\ \bibnamefont {Glenzer}},\ and\ \bibinfo
  {author} {\bibfnamefont {R.}~\bibnamefont {Redmer}},\ }\href
  {https://doi.org/10.1063/1.5017889} {\bibfield  {journal} {\bibinfo
  {journal} {Physics of Plasmas}\ }\textbf {\bibinfo {volume} {25}},\ \bibinfo
  {pages} {056901} (\bibinfo {year} {2018})}\BibitemShut {NoStop}%
\bibitem [{\citenamefont {Holst}\ \emph {et~al.}(2008)\citenamefont {Holst},
  \citenamefont {Redmer},\ and\ \citenamefont {Desjarlais}}]{Holst2008}%
  \BibitemOpen
  \bibfield  {author} {\bibinfo {author} {\bibfnamefont {B.}~\bibnamefont
  {Holst}}, \bibinfo {author} {\bibfnamefont {R.}~\bibnamefont {Redmer}},\ and\
  \bibinfo {author} {\bibfnamefont {M.~P.}\ \bibnamefont {Desjarlais}},\
  }\href@noop {} {\bibfield  {journal} {\bibinfo  {journal} {Phys. Rev. B}\
  }\textbf {\bibinfo {volume} {77}},\ \bibinfo {pages} {184201} (\bibinfo
  {year} {2008})}\BibitemShut {NoStop}%
\bibitem [{\citenamefont {Preising}\ \emph {et~al.}(2018)\citenamefont
  {Preising}, \citenamefont {Lorenzen}, \citenamefont {Becker}, \citenamefont
  {Redmer}, \citenamefont {Knudson},\ and\ \citenamefont
  {Desjarlais}}]{Preising2018}%
  \BibitemOpen
  \bibfield  {author} {\bibinfo {author} {\bibfnamefont {M.}~\bibnamefont
  {Preising}}, \bibinfo {author} {\bibfnamefont {W.}~\bibnamefont {Lorenzen}},
  \bibinfo {author} {\bibfnamefont {A.}~\bibnamefont {Becker}}, \bibinfo
  {author} {\bibfnamefont {R.}~\bibnamefont {Redmer}}, \bibinfo {author}
  {\bibfnamefont {M.~D.}\ \bibnamefont {Knudson}},\ and\ \bibinfo {author}
  {\bibfnamefont {M.~P.}\ \bibnamefont {Desjarlais}},\ }\href@noop {}
  {\bibfield  {journal} {\bibinfo  {journal} {Phys. Plasmas}\ }\textbf
  {\bibinfo {volume} {25}},\ \bibinfo {pages} {012706} (\bibinfo {year}
  {2018})}\BibitemShut {NoStop}%
\bibitem [{\citenamefont {Kubo}(1957)}]{kubo1957statisticalmechanical}%
  \BibitemOpen
  \bibfield  {author} {\bibinfo {author} {\bibfnamefont {R.}~\bibnamefont
  {Kubo}},\ }\href {https://doi.org/10.1143/JPSJ.12.570} {\bibfield  {journal}
  {\bibinfo  {journal} {J. Phys. Soc. Jpn.}\ }\textbf {\bibinfo {volume}
  {12}},\ \bibinfo {pages} {570} (\bibinfo {year} {1957})}\BibitemShut
  {NoStop}%
\bibitem [{\citenamefont {Mazevet}\ \emph {et~al.}(2010)\citenamefont
  {Mazevet}, \citenamefont {Torrent}, \citenamefont {Recoules},\ and\
  \citenamefont {Jollet}}]{mazevet2010calculations}%
  \BibitemOpen
  \bibfield  {author} {\bibinfo {author} {\bibfnamefont {S.}~\bibnamefont
  {Mazevet}}, \bibinfo {author} {\bibfnamefont {M.}~\bibnamefont {Torrent}},
  \bibinfo {author} {\bibfnamefont {V.}~\bibnamefont {Recoules}},\ and\
  \bibinfo {author} {\bibfnamefont {F.}~\bibnamefont {Jollet}},\ }\href
  {https://doi.org/10.1016/j.hedp.2009.06.004} {\bibfield  {journal} {\bibinfo
  {journal} {High Energy Density Physics}\ }\textbf {\bibinfo {volume} {6}},\
  \bibinfo {pages} {84} (\bibinfo {year} {2010})}\BibitemShut {NoStop}%
\bibitem [{\citenamefont {Holst}\ \emph {et~al.}(2011)\citenamefont {Holst},
  \citenamefont {French},\ and\ \citenamefont {Redmer}}]{holst2011electronic}%
  \BibitemOpen
  \bibfield  {author} {\bibinfo {author} {\bibfnamefont {B.}~\bibnamefont
  {Holst}}, \bibinfo {author} {\bibfnamefont {M.}~\bibnamefont {French}},\ and\
  \bibinfo {author} {\bibfnamefont {R.}~\bibnamefont {Redmer}},\ }\href
  {https://doi.org/10.1103/PhysRevB.83.235120} {\bibfield  {journal} {\bibinfo
  {journal} {Physical Review B}\ }\textbf {\bibinfo {volume} {83}},\ \bibinfo
  {pages} {235120} (\bibinfo {year} {2011})}\BibitemShut {NoStop}%
\bibitem [{\citenamefont {Desjarlais}\ \emph {et~al.}(2002)\citenamefont
  {Desjarlais}, \citenamefont {Kress},\ and\ \citenamefont
  {Collins}}]{desjarlais2002electrical}%
  \BibitemOpen
  \bibfield  {author} {\bibinfo {author} {\bibfnamefont {M.~P.}\ \bibnamefont
  {Desjarlais}}, \bibinfo {author} {\bibfnamefont {J.~D.}\ \bibnamefont
  {Kress}},\ and\ \bibinfo {author} {\bibfnamefont {L.~A.}\ \bibnamefont
  {Collins}},\ }\href {https://doi.org/10.1103/PhysRevE.66.025401} {\bibfield
  {journal} {\bibinfo  {journal} {Physical Review E}\ }\textbf {\bibinfo
  {volume} {66}},\ \bibinfo {pages} {025401} (\bibinfo {year}
  {2002})}\BibitemShut {NoStop}%
\bibitem [{\citenamefont {Cytter}\ \emph {et~al.}(2018)\citenamefont {Cytter},
  \citenamefont {Rabani}, \citenamefont {Neuhauser},\ and\ \citenamefont
  {Baer}}]{cytter2018stochastic}%
  \BibitemOpen
  \bibfield  {author} {\bibinfo {author} {\bibfnamefont {Y.}~\bibnamefont
  {Cytter}}, \bibinfo {author} {\bibfnamefont {E.}~\bibnamefont {Rabani}},
  \bibinfo {author} {\bibfnamefont {D.}~\bibnamefont {Neuhauser}},\ and\
  \bibinfo {author} {\bibfnamefont {R.}~\bibnamefont {Baer}},\ }\href@noop {}
  {\bibfield  {journal} {\bibinfo  {journal} {Phys. Rev. B}\ }\textbf {\bibinfo
  {volume} {97}},\ \bibinfo {pages} {115207} (\bibinfo {year}
  {2018})}\BibitemShut {NoStop}%
\bibitem [{\citenamefont {Baer}\ \emph {et~al.}(2013)\citenamefont {Baer},
  \citenamefont {Neuhauser},\ and\ \citenamefont {Rabani}}]{Baer2013}%
  \BibitemOpen
  \bibfield  {author} {\bibinfo {author} {\bibfnamefont {R.}~\bibnamefont
  {Baer}}, \bibinfo {author} {\bibfnamefont {D.}~\bibnamefont {Neuhauser}},\
  and\ \bibinfo {author} {\bibfnamefont {E.}~\bibnamefont {Rabani}},\ }\href
  {https://doi.org/10.1103/PhysRevLett.111.106402} {\bibfield  {journal}
  {\bibinfo  {journal} {Phys. Rev. Lett.}\ }\textbf {\bibinfo {volume} {111}},\
  \bibinfo {pages} {106402} (\bibinfo {year} {2013})}\BibitemShut {NoStop}%
\bibitem [{\citenamefont {Neuhauser}\ \emph {et~al.}(2014)\citenamefont
  {Neuhauser}, \citenamefont {Baer},\ and\ \citenamefont
  {Rabani}}]{Neuhauser2014a}%
  \BibitemOpen
  \bibfield  {author} {\bibinfo {author} {\bibfnamefont {D.}~\bibnamefont
  {Neuhauser}}, \bibinfo {author} {\bibfnamefont {R.}~\bibnamefont {Baer}},\
  and\ \bibinfo {author} {\bibfnamefont {E.}~\bibnamefont {Rabani}},\
  }\href@noop {} {\bibfield  {journal} {\bibinfo  {journal} {J. Chem. Phys.}\
  }\textbf {\bibinfo {volume} {141}},\ \bibinfo {pages} {041102} (\bibinfo
  {year} {2014})}\BibitemShut {NoStop}%
\bibitem [{\citenamefont {Cytter}\ \emph {et~al.}(2014)\citenamefont {Cytter},
  \citenamefont {Neuhauser},\ and\ \citenamefont
  {Baer}}]{cytter2014metropolis}%
  \BibitemOpen
  \bibfield  {author} {\bibinfo {author} {\bibfnamefont {Y.}~\bibnamefont
  {Cytter}}, \bibinfo {author} {\bibfnamefont {D.}~\bibnamefont {Neuhauser}},\
  and\ \bibinfo {author} {\bibfnamefont {R.}~\bibnamefont {Baer}},\ }\href
  {https://doi.org/10.1021/ct500450w} {\bibfield  {journal} {\bibinfo
  {journal} {J. Chem. Theory Comput.}\ }\textbf {\bibinfo {volume} {10}},\
  \bibinfo {pages} {4317} (\bibinfo {year} {2014})}\BibitemShut {NoStop}%
\bibitem [{\citenamefont {Fabian}\ \emph {et~al.}(2018)\citenamefont {Fabian},
  \citenamefont {Shpiro}, \citenamefont {Rabani}, \citenamefont {Neuhauser},\
  and\ \citenamefont {Baer}}]{fabian2018stochastic}%
  \BibitemOpen
  \bibfield  {author} {\bibinfo {author} {\bibfnamefont {M.~D.}\ \bibnamefont
  {Fabian}}, \bibinfo {author} {\bibfnamefont {B.}~\bibnamefont {Shpiro}},
  \bibinfo {author} {\bibfnamefont {E.}~\bibnamefont {Rabani}}, \bibinfo
  {author} {\bibfnamefont {D.}~\bibnamefont {Neuhauser}},\ and\ \bibinfo
  {author} {\bibfnamefont {R.}~\bibnamefont {Baer}},\ }\href
  {https://doi.org/10.1002/wcms.1412} {\bibfield  {journal} {\bibinfo
  {journal} {Wiley Interdisciplinary Reviews: Computational Molecular Science}\
  }\textbf {\bibinfo {volume} {10.1002/wcms.1412}},\ \bibinfo {pages} {e1412}
  (\bibinfo {year} {2018})}\BibitemShut {NoStop}%
\bibitem [{\citenamefont {Chen}\ \emph {et~al.}(2019)\citenamefont {Chen},
  \citenamefont {Baer}, \citenamefont {Neuhauser},\ and\ \citenamefont
  {Rabani}}]{chen2019overlapped}%
  \BibitemOpen
  \bibfield  {author} {\bibinfo {author} {\bibfnamefont {M.}~\bibnamefont
  {Chen}}, \bibinfo {author} {\bibfnamefont {R.}~\bibnamefont {Baer}}, \bibinfo
  {author} {\bibfnamefont {D.}~\bibnamefont {Neuhauser}},\ and\ \bibinfo
  {author} {\bibfnamefont {E.}~\bibnamefont {Rabani}},\ }\href
  {https://doi.org/10.1063/1.5064472} {\bibfield  {journal} {\bibinfo
  {journal} {J. Chem. Phys.}\ }\textbf {\bibinfo {volume} {150}},\ \bibinfo
  {pages} {034106} (\bibinfo {year} {2019})}\BibitemShut {NoStop}%
\bibitem [{\citenamefont {Gao}\ \emph {et~al.}(2015)\citenamefont {Gao},
  \citenamefont {Neuhauser}, \citenamefont {Baer},\ and\ \citenamefont
  {Rabani}}]{Gao2015}%
  \BibitemOpen
  \bibfield  {author} {\bibinfo {author} {\bibfnamefont {Y.}~\bibnamefont
  {Gao}}, \bibinfo {author} {\bibfnamefont {D.}~\bibnamefont {Neuhauser}},
  \bibinfo {author} {\bibfnamefont {R.}~\bibnamefont {Baer}},\ and\ \bibinfo
  {author} {\bibfnamefont {E.}~\bibnamefont {Rabani}},\ }\href@noop {}
  {\bibfield  {journal} {\bibinfo  {journal} {J. Chem. Phys.}\ }\textbf
  {\bibinfo {volume} {142}},\ \bibinfo {pages} {034106} (\bibinfo {year}
  {2015})}\BibitemShut {NoStop}%
\bibitem [{\citenamefont {Neuhauser}\ \emph {et~al.}(2017)\citenamefont
  {Neuhauser}, \citenamefont {Baer},\ and\ \citenamefont
  {Zgid}}]{Neuhauser2017}%
  \BibitemOpen
  \bibfield  {author} {\bibinfo {author} {\bibfnamefont {D.}~\bibnamefont
  {Neuhauser}}, \bibinfo {author} {\bibfnamefont {R.}~\bibnamefont {Baer}},\
  and\ \bibinfo {author} {\bibfnamefont {D.}~\bibnamefont {Zgid}},\ }\href@noop
  {} {\bibfield  {journal} {\bibinfo  {journal} {J. Chem. Theory Comput.}\
  }\textbf {\bibinfo {volume} {13}},\ \bibinfo {pages} {5396} (\bibinfo {year}
  {2017})}\BibitemShut {NoStop}%
\bibitem [{\citenamefont {Hernandez}\ \emph {et~al.}(2018)\citenamefont
  {Hernandez}, \citenamefont {Xia}, \citenamefont {Vl{\v{c}}ek}, \citenamefont
  {Boutelle}, \citenamefont {Baer}, \citenamefont {Rabani},\ and\ \citenamefont
  {Neuhauser}}]{hernandez2018first}%
  \BibitemOpen
  \bibfield  {author} {\bibinfo {author} {\bibfnamefont {S.}~\bibnamefont
  {Hernandez}}, \bibinfo {author} {\bibfnamefont {Y.}~\bibnamefont {Xia}},
  \bibinfo {author} {\bibfnamefont {V.}~\bibnamefont {Vl{\v{c}}ek}}, \bibinfo
  {author} {\bibfnamefont {R.}~\bibnamefont {Boutelle}}, \bibinfo {author}
  {\bibfnamefont {R.}~\bibnamefont {Baer}}, \bibinfo {author} {\bibfnamefont
  {E.}~\bibnamefont {Rabani}},\ and\ \bibinfo {author} {\bibfnamefont
  {D.}~\bibnamefont {Neuhauser}},\ }\href@noop {} {\bibfield  {journal}
  {\bibinfo  {journal} {Mol. Phys.}\ }\textbf {\bibinfo {volume} {116}},\
  \bibinfo {pages} {2506} (\bibinfo {year} {2018})}\BibitemShut {NoStop}%
\bibitem [{\citenamefont {Takeshita}\ \emph {et~al.}(2017)\citenamefont
  {Takeshita}, \citenamefont {de~Jong}, \citenamefont {Neuhauser},
  \citenamefont {Baer},\ and\ \citenamefont
  {Rabani}}]{takeshita2017stochastic}%
  \BibitemOpen
  \bibfield  {author} {\bibinfo {author} {\bibfnamefont {T.~Y.}\ \bibnamefont
  {Takeshita}}, \bibinfo {author} {\bibfnamefont {W.~A.}\ \bibnamefont
  {de~Jong}}, \bibinfo {author} {\bibfnamefont {D.}~\bibnamefont {Neuhauser}},
  \bibinfo {author} {\bibfnamefont {R.}~\bibnamefont {Baer}},\ and\ \bibinfo
  {author} {\bibfnamefont {E.}~\bibnamefont {Rabani}},\ }\href
  {https://doi.org/10.1021/acs.jctc.7b00343} {\bibfield  {journal} {\bibinfo
  {journal} {J. Chem. Theory Comput.}\ }\textbf {\bibinfo {volume} {13}},\
  \bibinfo {pages} {4605} (\bibinfo {year} {2017})},\ \Eprint
  {https://arxiv.org/abs/http://dx.doi.org/10.1021/acs.jctc.7b00343}
  {http://dx.doi.org/10.1021/acs.jctc.7b00343} \BibitemShut {NoStop}%
\bibitem [{\citenamefont {Wang}(1994)}]{Wang1994g}%
  \BibitemOpen
  \bibfield  {author} {\bibinfo {author} {\bibfnamefont {L.-W.}\ \bibnamefont
  {Wang}},\ }\href@noop {} {\bibfield  {journal} {\bibinfo  {journal} {Phys.
  Rev. B}\ }\textbf {\bibinfo {volume} {49}},\ \bibinfo {pages} {10154}
  (\bibinfo {year} {1994})}\BibitemShut {NoStop}%
\bibitem [{\citenamefont {Baer}\ \emph {et~al.}(2004)\citenamefont {Baer},
  \citenamefont {Seideman}, \citenamefont {Ilani},\ and\ \citenamefont
  {Neuhauser}}]{Baer2004c}%
  \BibitemOpen
  \bibfield  {author} {\bibinfo {author} {\bibfnamefont {R.}~\bibnamefont
  {Baer}}, \bibinfo {author} {\bibfnamefont {T.}~\bibnamefont {Seideman}},
  \bibinfo {author} {\bibfnamefont {S.}~\bibnamefont {Ilani}},\ and\ \bibinfo
  {author} {\bibfnamefont {D.}~\bibnamefont {Neuhauser}},\ }\href@noop {}
  {\bibfield  {journal} {\bibinfo  {journal} {J. Chem. Phys.}\ }\textbf
  {\bibinfo {volume} {120}},\ \bibinfo {pages} {3387} (\bibinfo {year}
  {2004})}\BibitemShut {NoStop}%
\bibitem [{\citenamefont {Iitaka}\ \emph {et~al.}(1997)\citenamefont {Iitaka},
  \citenamefont {Nomura}, \citenamefont {Hirayama}, \citenamefont {Zhao},
  \citenamefont {Aoyagi},\ and\ \citenamefont {Sugano}}]{Iitaka1997}%
  \BibitemOpen
  \bibfield  {author} {\bibinfo {author} {\bibfnamefont {T.}~\bibnamefont
  {Iitaka}}, \bibinfo {author} {\bibfnamefont {S.}~\bibnamefont {Nomura}},
  \bibinfo {author} {\bibfnamefont {H.}~\bibnamefont {Hirayama}}, \bibinfo
  {author} {\bibfnamefont {X.~W.}\ \bibnamefont {Zhao}}, \bibinfo {author}
  {\bibfnamefont {Y.}~\bibnamefont {Aoyagi}},\ and\ \bibinfo {author}
  {\bibfnamefont {T.}~\bibnamefont {Sugano}},\ }\href@noop {} {\bibfield
  {journal} {\bibinfo  {journal} {Phys. Rev. E}\ }\textbf {\bibinfo {volume}
  {56}},\ \bibinfo {pages} {1222} (\bibinfo {year} {1997})}\BibitemShut
  {NoStop}%
\bibitem [{\citenamefont {Kubo}(1966)}]{kubo1966thefluctuationdissipation}%
  \BibitemOpen
  \bibfield  {author} {\bibinfo {author} {\bibfnamefont {R.}~\bibnamefont
  {Kubo}},\ }\href@noop {} {\bibfield  {journal} {\bibinfo  {journal} {Rep.
  Prog. Phys.}\ }\textbf {\bibinfo {volume} {29}},\ \bibinfo {pages} {255}
  (\bibinfo {year} {1966})}\BibitemShut {NoStop}%
\bibitem [{\citenamefont {Greenwood}(1958)}]{greenwood1958theboltzmann}%
  \BibitemOpen
  \bibfield  {author} {\bibinfo {author} {\bibfnamefont {D.~A.}\ \bibnamefont
  {Greenwood}},\ }\href {https://doi.org/10.1088/0370-1328/71/4/306} {\bibfield
   {journal} {\bibinfo  {journal} {Proc. Phys. Soc.}\ }\textbf {\bibinfo
  {volume} {71}},\ \bibinfo {pages} {585} (\bibinfo {year} {1958})}\BibitemShut
  {NoStop}%
\bibitem [{\citenamefont {Hutchinson}(1990)}]{Hutchinson1990}%
  \BibitemOpen
  \bibfield  {author} {\bibinfo {author} {\bibfnamefont {M.~F.}\ \bibnamefont
  {Hutchinson}},\ }\href@noop {} {\bibfield  {journal} {\bibinfo  {journal}
  {Commun Stat Simul Comput.}\ }\textbf {\bibinfo {volume} {19}},\ \bibinfo
  {pages} {433} (\bibinfo {year} {1990})}\BibitemShut {NoStop}%
\bibitem [{\citenamefont {Perdew}\ and\ \citenamefont
  {Wang}(1992)}]{Perdew1992a}%
  \BibitemOpen
  \bibfield  {author} {\bibinfo {author} {\bibfnamefont {J.}~\bibnamefont
  {Perdew}}\ and\ \bibinfo {author} {\bibfnamefont {Y.}~\bibnamefont {Wang}},\
  }\href@noop {} {\bibfield  {journal} {\bibinfo  {journal} {Phys. Rev. B}\
  }\textbf {\bibinfo {volume} {45}},\ \bibinfo {pages} {13244} (\bibinfo {year}
  {1992})}\BibitemShut {NoStop}%
\bibitem [{\citenamefont {Troullier}\ and\ \citenamefont
  {Martins}(1991)}]{Troullier1991}%
  \BibitemOpen
  \bibfield  {author} {\bibinfo {author} {\bibfnamefont {N.}~\bibnamefont
  {Troullier}}\ and\ \bibinfo {author} {\bibfnamefont {J.~L.}\ \bibnamefont
  {Martins}},\ }\href@noop {} {\bibfield  {journal} {\bibinfo  {journal} {Phys.
  Rev. B}\ }\textbf {\bibinfo {volume} {43}},\ \bibinfo {pages} {1993}
  (\bibinfo {year} {1991})}\BibitemShut {NoStop}%
\bibitem [{\citenamefont {Kleinman}\ and\ \citenamefont
  {Bylander}(1982)}]{Kleinman1982}%
  \BibitemOpen
  \bibfield  {author} {\bibinfo {author} {\bibfnamefont {L.}~\bibnamefont
  {Kleinman}}\ and\ \bibinfo {author} {\bibfnamefont {D.}~\bibnamefont
  {Bylander}},\ }\href@noop {} {\bibfield  {journal} {\bibinfo  {journal}
  {Phys. Rev. Lett.}\ }\textbf {\bibinfo {volume} {48}},\ \bibinfo {pages}
  {1425} (\bibinfo {year} {1982})}\BibitemShut {NoStop}%
\bibitem [{\citenamefont {Giannozzi}\ \emph {et~al.}(2009)\citenamefont
  {Giannozzi}, \citenamefont {Baroni} \emph {et~al.}}]{Giannozzi2009}%
  \BibitemOpen
  \bibfield  {author} {\bibinfo {author} {\bibfnamefont {P.}~\bibnamefont
  {Giannozzi}}, \bibinfo {author} {\bibfnamefont {S.}~\bibnamefont {Baroni}},
  \emph {et~al.},\ }\href {http://stacks.iop.org/0953-8984/21/i=39/a=395502}
  {\bibfield  {journal} {\bibinfo  {journal} {J. Phys.: Condens. Matter}\
  }\textbf {\bibinfo {volume} {21}},\ \bibinfo {pages} {395502} (\bibinfo
  {year} {2009})}\BibitemShut {NoStop}%
\bibitem [{\citenamefont {Calder{\'i}n}\ \emph {et~al.}(2017)\citenamefont
  {Calder{\'i}n}, \citenamefont {Karasiev},\ and\ \citenamefont
  {Trickey}}]{calderin2017kubotextendashgreenwood}%
  \BibitemOpen
  \bibfield  {author} {\bibinfo {author} {\bibfnamefont {L.}~\bibnamefont
  {Calder{\'i}n}}, \bibinfo {author} {\bibfnamefont {V.~V.}\ \bibnamefont
  {Karasiev}},\ and\ \bibinfo {author} {\bibfnamefont {S.~B.}\ \bibnamefont
  {Trickey}},\ }\href {https://doi.org/10.1016/j.cpc.2017.08.008} {\bibfield
  {journal} {\bibinfo  {journal} {Computer Physics Communications}\ }\textbf
  {\bibinfo {volume} {221}},\ \bibinfo {pages} {118} (\bibinfo {year}
  {2017})}\BibitemShut {NoStop}%
\bibitem [{\citenamefont {Perdew}\ \emph {et~al.}(1996)\citenamefont {Perdew},
  \citenamefont {Burke},\ and\ \citenamefont {Ernzerhof}}]{Perdew1996}%
  \BibitemOpen
  \bibfield  {author} {\bibinfo {author} {\bibfnamefont {J.~P.}\ \bibnamefont
  {Perdew}}, \bibinfo {author} {\bibfnamefont {K.}~\bibnamefont {Burke}},\ and\
  \bibinfo {author} {\bibfnamefont {M.}~\bibnamefont {Ernzerhof}},\ }\href@noop
  {} {\bibfield  {journal} {\bibinfo  {journal} {Phys. Rev. Lett.}\ }\textbf
  {\bibinfo {volume} {77}},\ \bibinfo {pages} {3865} (\bibinfo {year}
  {1996})}\BibitemShut {NoStop}%
\bibitem [{\citenamefont {Hafner}(2007)}]{Hafner2007}%
  \BibitemOpen
  \bibfield  {author} {\bibinfo {author} {\bibfnamefont {J.}~\bibnamefont
  {Hafner}},\ }\href@noop {} {\bibfield  {journal} {\bibinfo  {journal}
  {Comput. Phys. Commun.}\ }\textbf {\bibinfo {volume} {177}},\ \bibinfo
  {pages} {6} (\bibinfo {year} {2007})}\BibitemShut {NoStop}%
\bibitem [{\citenamefont {Kresse}\ and\ \citenamefont
  {Furthmuller}(1996)}]{Kresse1996}%
  \BibitemOpen
  \bibfield  {author} {\bibinfo {author} {\bibfnamefont {G.}~\bibnamefont
  {Kresse}}\ and\ \bibinfo {author} {\bibfnamefont {J.}~\bibnamefont
  {Furthmuller}},\ }\href@noop {} {\bibfield  {journal} {\bibinfo  {journal}
  {Phys. Rev. B}\ }\textbf {\bibinfo {volume} {54}},\ \bibinfo {pages} {11169}
  (\bibinfo {year} {1996})}\BibitemShut {NoStop}%
\bibitem [{\citenamefont {Kuhn}(1925)}]{kuhn1925aoeber}%
  \BibitemOpen
  \bibfield  {author} {\bibinfo {author} {\bibfnamefont {W.}~\bibnamefont
  {Kuhn}},\ }\href@noop {} {\bibfield  {journal} {\bibinfo  {journal} {Z.
  Angew. Phys.}\ }\textbf {\bibinfo {volume} {33}},\ \bibinfo {pages} {408}
  (\bibinfo {year} {1925})}\BibitemShut {NoStop}%
\bibitem [{\citenamefont {Reiche}\ and\ \citenamefont
  {Thomas}(1925)}]{reiche1925uberdie}%
  \BibitemOpen
  \bibfield  {author} {\bibinfo {author} {\bibfnamefont {F.}~\bibnamefont
  {Reiche}}\ and\ \bibinfo {author} {\bibfnamefont {W.}~\bibnamefont
  {Thomas}},\ }\href {https://doi.org/10.1007/BF01328494} {\bibfield  {journal}
  {\bibinfo  {journal} {Zeitschrift f{\"u}r Physik}\ }\textbf {\bibinfo
  {volume} {34}},\ \bibinfo {pages} {510} (\bibinfo {year} {1925})}\BibitemShut
  {NoStop}%
\bibitem [{\citenamefont {Kosloff}(1988)}]{Kosloff1988}%
  \BibitemOpen
  \bibfield  {author} {\bibinfo {author} {\bibfnamefont {R.}~\bibnamefont
  {Kosloff}},\ }\href@noop {} {\bibfield  {journal} {\bibinfo  {journal} {J.
  Phys. Chem.}\ }\textbf {\bibinfo {volume} {92}},\ \bibinfo {pages} {2087}
  (\bibinfo {year} {1988})}\BibitemShut {NoStop}%
\bibitem [{\citenamefont {Goedecker}\ and\ \citenamefont
  {Colombo}(1994)}]{Goedecker1994}%
  \BibitemOpen
  \bibfield  {author} {\bibinfo {author} {\bibfnamefont {S.}~\bibnamefont
  {Goedecker}}\ and\ \bibinfo {author} {\bibfnamefont {L.}~\bibnamefont
  {Colombo}},\ }\href@noop {} {\bibfield  {journal} {\bibinfo  {journal} {Phys.
  Rev. Lett.}\ }\textbf {\bibinfo {volume} {73}},\ \bibinfo {pages} {122}
  (\bibinfo {year} {1994})}\BibitemShut {NoStop}%
\bibitem [{\citenamefont {Huang}\ \emph {et~al.}(1995)\citenamefont {Huang},
  \citenamefont {Kouri},\ and\ \citenamefont {Hoffman}}]{Huang1995}%
  \BibitemOpen
  \bibfield  {author} {\bibinfo {author} {\bibfnamefont {Y.~H.}\ \bibnamefont
  {Huang}}, \bibinfo {author} {\bibfnamefont {D.~J.}\ \bibnamefont {Kouri}},\
  and\ \bibinfo {author} {\bibfnamefont {D.~K.}\ \bibnamefont {Hoffman}},\
  }\href@noop {} {\bibfield  {journal} {\bibinfo  {journal} {Chem. Phys.
  Lett.}\ }\textbf {\bibinfo {volume} {243}},\ \bibinfo {pages} {367} (\bibinfo
  {year} {1995})}\BibitemShut {NoStop}%
\bibitem [{\citenamefont {Baer}\ and\ \citenamefont
  {Head-Gordon}(1997{\natexlab{a}})}]{Baer1997a}%
  \BibitemOpen
  \bibfield  {author} {\bibinfo {author} {\bibfnamefont {R.}~\bibnamefont
  {Baer}}\ and\ \bibinfo {author} {\bibfnamefont {M.}~\bibnamefont
  {Head-Gordon}},\ }\href@noop {} {\bibfield  {journal} {\bibinfo  {journal}
  {J. Chem. Phys.}\ }\textbf {\bibinfo {volume} {107}},\ \bibinfo {pages}
  {10003} (\bibinfo {year} {1997}{\natexlab{a}})}\BibitemShut {NoStop}%
\bibitem [{Note1()}]{Note1}%
  \BibitemOpen
  \bibinfo {note} {The Chebyshev recursion is $\phi _{m+1}=2\protect
  \mathaccentV {hat}05E{h}_{N}\phi _{m}-\phi _{m-1}$, with $\phi _{0}=\varphi $
  and $\phi _{1}=\protect \mathaccentV {hat}05E{h}_{N}\phi _{0}$, where
  $\protect \mathaccentV {hat}05E{h}_{N}=\protect \frac {\protect \mathaccentV
  {hat}05E{h}-\protect \mathaccentV {bar}016{E}}{\Delta E}$, $\protect
  \mathaccentV {bar}016{E}=\protect \frac {1}{2}\left (E_{max}+E_{min}\right )$
  and $\Delta E=\protect \frac {1}{2}\left (E_{max}-E_{min}\right
  )$.}\BibitemShut {Stop}%
\bibitem [{\citenamefont {Baer}\ and\ \citenamefont
  {Head-Gordon}(1997{\natexlab{b}})}]{baer1997sparsity}%
  \BibitemOpen
  \bibfield  {author} {\bibinfo {author} {\bibfnamefont {R.}~\bibnamefont
  {Baer}}\ and\ \bibinfo {author} {\bibfnamefont {M.}~\bibnamefont
  {Head-Gordon}},\ }\href {https://doi.org/10.1103/PhysRevLett.79.3962}
  {\bibfield  {journal} {\bibinfo  {journal} {Phys. Rev. Lett.}\ }\textbf
  {\bibinfo {volume} {79}},\ \bibinfo {pages} {3962} (\bibinfo {year}
  {1997}{\natexlab{b}})}\BibitemShut {NoStop}%
\bibitem [{\citenamefont {Jaramillo-Botero}\ \emph {et~al.}(2010)\citenamefont
  {Jaramillo-Botero}, \citenamefont {Su}, \citenamefont {Qi},\ and\
  \citenamefont {Goddard}}]{JaramilloBotero2010}%
  \BibitemOpen
  \bibfield  {author} {\bibinfo {author} {\bibfnamefont {A.}~\bibnamefont
  {Jaramillo-Botero}}, \bibinfo {author} {\bibfnamefont {J.}~\bibnamefont
  {Su}}, \bibinfo {author} {\bibfnamefont {A.}~\bibnamefont {Qi}},\ and\
  \bibinfo {author} {\bibfnamefont {W.~A.}\ \bibnamefont {Goddard}},\ }\href
  {https://doi.org/10.1002/jcc.21637} {\bibfield  {journal} {\bibinfo
  {journal} {J. Comput. Chem.}\ }\textbf {\bibinfo {volume} {32}},\ \bibinfo
  {pages} {497} (\bibinfo {year} {2010})}\BibitemShut {NoStop}%
\bibitem [{\citenamefont {Plimpton}(1995)}]{plimpton1995fastparallel}%
  \BibitemOpen
  \bibfield  {author} {\bibinfo {author} {\bibfnamefont {S.}~\bibnamefont
  {Plimpton}},\ }\href {https://doi.org/10.1006/jcph.1995.1039} {\bibfield
  {journal} {\bibinfo  {journal} {Journal of Computational Physics}\ }\textbf
  {\bibinfo {volume} {117}},\ \bibinfo {pages} {1} (\bibinfo {year}
  {1995})}\BibitemShut {NoStop}%
\bibitem [{\citenamefont {Kim}\ \emph {et~al.}(2011)\citenamefont {Kim},
  \citenamefont {Su},\ and\ \citenamefont {Goddard}}]{kim2011hightemperature}%
  \BibitemOpen
  \bibfield  {author} {\bibinfo {author} {\bibfnamefont {H.}~\bibnamefont
  {Kim}}, \bibinfo {author} {\bibfnamefont {J.~T.}\ \bibnamefont {Su}},\ and\
  \bibinfo {author} {\bibfnamefont {W.~A.}\ \bibnamefont {Goddard}},\ }\href
  {https://doi.org/10.1073/pnas.1110322108} {\bibfield  {journal} {\bibinfo
  {journal} {PNAS}\ }\textbf {\bibinfo {volume} {108}},\ \bibinfo {pages}
  {15101} (\bibinfo {year} {2011})}\BibitemShut {NoStop}%
\bibitem [{\citenamefont {Su}\ and\ \citenamefont
  {Goddard}(2007)}]{su2007excited}%
  \BibitemOpen
  \bibfield  {author} {\bibinfo {author} {\bibfnamefont {J.~T.}\ \bibnamefont
  {Su}}\ and\ \bibinfo {author} {\bibfnamefont {W.~A.}\ \bibnamefont
  {Goddard}},\ }\href {https://doi.org/10.1103/PhysRevLett.99.185003}
  {\bibfield  {journal} {\bibinfo  {journal} {Phys. Rev. Lett.}\ }\textbf
  {\bibinfo {volume} {99}},\ \bibinfo {pages} {185003} (\bibinfo {year}
  {2007})}\BibitemShut {NoStop}%
\bibitem [{\citenamefont {Lorenzen}\ \emph {et~al.}(2009)\citenamefont
  {Lorenzen}, \citenamefont {Holst},\ and\ \citenamefont
  {Redmer}}]{lorenzen2009demixing}%
  \BibitemOpen
  \bibfield  {author} {\bibinfo {author} {\bibfnamefont {W.}~\bibnamefont
  {Lorenzen}}, \bibinfo {author} {\bibfnamefont {B.}~\bibnamefont {Holst}},\
  and\ \bibinfo {author} {\bibfnamefont {R.}~\bibnamefont {Redmer}},\ }\href
  {https://doi.org/10.1103/PhysRevLett.102.115701} {\bibfield  {journal}
  {\bibinfo  {journal} {Physical Review Letters}\ }\textbf {\bibinfo {volume}
  {102}},\ \bibinfo {pages} {115701} (\bibinfo {year} {2009})}\BibitemShut
  {NoStop}%
\end{thebibliography}

\end{document}